\documentclass[%
reprint,
nofootinbib,
 amsmath,amssymb,
aps,
pra,
]{revtex4-2}

\usepackage{float}

\usepackage{graphicx}
\usepackage[final]{hyperref}
\usepackage{dcolumn}
\usepackage{bm}
\usepackage{gensymb}
\usepackage{amsmath}
\usepackage{xcolor}

\newcommand{\dd}{\mathrm{d}}
\newcommand{\defeq}{\stackrel{\mathrm{def}}{=}}
\newcommand{\dB}{\text{dB}}

\newcommand{\tin}{\text{in}}
\newcommand{\out}{\text{out}}
\newcommand{\sig}{\text{s}}
\newcommand{\pump}{\text{p}}

\usepackage{CJKutf8}    

\begin{document}


\title{Pump-efficient Josephson parametric amplifiers with high saturation power}

\author{Nicholas M. Hougland}
  \thanks{These authors contributed equally to this work.}
  \affiliation{Department of Physics and Astronomy, University of Pittsburgh}
  \affiliation{Pittsburgh Quantum Institute, University of Pittsburgh}
\author{Zhuan Li (李專)}
  \thanks{These authors contributed equally to this work.}
  \affiliation{Department of Physics and Astronomy, University of Pittsburgh}
  \affiliation{Pittsburgh Quantum Institute, University of Pittsburgh}
\author{Ryan Kaufman}%
  \affiliation{Department of Physics and Astronomy, University of Pittsburgh}
  \affiliation{Pittsburgh Quantum Institute, University of Pittsburgh}
\author{Boris Mesits}%
  \affiliation{Department of Physics and Astronomy, University of Pittsburgh}
  \affiliation{Pittsburgh Quantum Institute, University of Pittsburgh}
\author{Roger S. K. Mong (蒙紹璣)}%
  \affiliation{Department of Physics and Astronomy, University of Pittsburgh}
  \affiliation{Pittsburgh Quantum Institute, University of Pittsburgh}
\author{Michael Hatridge}%
  \affiliation{Department of Physics and Astronomy, University of Pittsburgh}
  \affiliation{Pittsburgh Quantum Institute, University of Pittsburgh}
\author{David Pekker}%
  \affiliation{Department of Physics and Astronomy, University of Pittsburgh}
  \affiliation{Pittsburgh Quantum Institute, University of Pittsburgh}

\date{\today}

\begin{abstract}
Circuit QED based quantum information processing relies on low noise amplification for signal readout. In the realm of microwave superconducting circuits, this amplification is often achieved via Josephson parametric amplifiers (JPA). In the past, these amplifiers exhibited low power added efficiency (PAE), which is roughly the fraction of pump power that is converted to output signal power. This is increasingly relevant because recent attempts to build high saturation power amplifiers achieve this at the cost of very low PAE, which in turn puts a high heat load on the cryostat and limits the number of these devices that a dilution refrigerator can host. Here, we numerically investigate upper bounds on PAE. We focus on a class of parametric amplifiers that consists of a capacitor shunted by a nonlinear inductive block. We first set a benchmark for this class of amplifiers by considering nonlinear blocks described by an arbitrary polynomial current-phase relation. Next, we identify that it is important for amplifiers with inductive blocks composed of repeating elements to have monotonic current-phase relations for each element in order to avoid exciting high-frequency modes. Using this design rule, we propose two circuit implementations for repeating elements in JPA inductive blocks. Finally, we investigate polynomial amplifier chains. We find that while amplifiers with higher gain have a lower PAE, regardless of the gain there is considerable room to improve as compared to state of the art devices. For example, for a degenerate amplifier with a power gain of 20~dB, the PAE is $\sim0.1\%$ for typical JPAs, 37.9\% for our simpler circuit JPAs, 42.6\% for our more complex circuit JPAs, 63.3\% for our arbitrary polynomial amplifiers, and at least 98\% for our amplifier chains.
\end{abstract}

\begin{CJK*}{UTF8}{bsmi}
\maketitle
\end{CJK*}

\section{Introduction}

The Josephson parametric amplifier (JPA), which has been extensively studied for its low noise profile~\cite{Feldman75,Feldman80,Yurke88,JPAs07}, plays an important role in superconducting quantum computing. It uses the Josephson junction to amplify weak microwave signals with nearly quantum-limited added noise. Due to its ultralow-noise performance, the JPA is vital for qubit state readout~\cite{qubitMeasure,linReadout,HatridgeMeasurement,Kaufman2023} for applications such as quantum state tomography~\cite{MalletTomography}, and, crucially, quantum error correction~\cite{DiCarloSurfaceCode,WallraffSurfaceCode,Schoelkopf2023,Sivak2023}.

A critical characteristic of the JPA is its ability to add the minimum noise throughout the parametric amplification process with a power gain of around 20~dB~\cite{caves,YurkeParametricAmplification, JoseSPAandJPA}. 
These noise and gain characteristics make JPAs suitable as the first stage of an amplifier chain, with the gain requirement being the minimum value to saturate the excess noise of commercial cryogenic high electron mobility transistor (HEMT) amplifiers~\cite{JoseSPAandJPA, lownoisefactory}.
Parametric amplification happens when the circuit’s parameter (e.g., inductance) is varied periodically at a specific frequency. In a microwave-pumped JPA, the periodic variation is achieved by using the nonlinearity of the Josephson junction to couple in a strong pump wave. 
Specifically, the nonlinearity mixes different frequency components of microwave signals. In particular, we consider the three-wave mixing case, where the pump wave at frequency $\omega_{\text{p}}$ and the signal wave at frequency $\omega_{\text{s}}$ mix and generate an idler wave at frequency $\omega_\text{i} = \omega_{\text{p}}-\omega_{\text{s}}$.
The parametric process transfers energy from the pump wave to the signal and idler waves and thus amplifies the incoming signals~\cite{JPCOriginal,JPCOriginal2,Roy2016}.

Based on signal and idler frequencies, parametric amplifiers are classified into phase-sensitive and phase-preserving amplifiers~\cite{JoseSPAandJPA}. The phase-sensitive amplifier is also called the degenerate amplifier because its signal and idler frequencies are hosted in the same physical mode.
Since the pump frequency is exactly twice the mode center frequency in this case, the power gain of the degenerate amplifier is sensitive to the relative phase between the pump and signal waves near the center of the amplifier's mode frequency. In contrast, the phase-preserving amplifier, also known as the nondegenerate amplifier, has different signal and idler frequencies.  In this case, the amplitude and phase information of the signal wave is maintained during the amplification process.

Besides added noise and gain, input saturation power and pump efficiency are also essential characteristics of the performance of parametric amplifiers~\cite{IEEEdynamicrange,dynamicJJparamp,optSNAILamp,sivak2019,Kaufman2023}.
The saturation power, $P_{\text{sat}}$, is defined as the smallest signal power at which the gain varies from the target gain $G_\text{t}$ by more than $\pm 1$~dB.
As quantum computing systems scale up with more qubits, we need to process signals of various amplitudes. Thus, a large saturation power is desired since it enables us to read qubit states over a wide range of input powers without distortion. Older amplifier designs like Josephson parametric converters had relatively low saturation power.
Surprisingly, the saturation power of these amplifiers was not limited by pump depletion. Instead, these amplifiers used a small number of Josephson junctions and, therefore, relatively small signal powers caused them to leave the desired nonlinear regime~\cite{liu2020JRM}. The newer generation of amplifiers splits the signal amplitude across many more Josephson junctions, thus resulting in a much larger saturation power~\cite{optSNAILamp,Parker2022threewavemixing,Kaufman2023}.

The subject of this paper is pump efficiency, which characterizes the capacity of an amplifier to convert input power from the pump to output signal power without distorting the signal. Specifically, we use the power added efficiency (PAE) as given on page~597 in Ref.~\cite{Pozar}, which is defined as
\begin{align}
 \eta_{\text{PAE}}(A_\text{s})  = \frac{P_{\text{out}}-P_{\text{in}}}{P_{\text{pump}}}=\frac{(G(A_\text{s})-1) \omega_\text{s}^2 A_\text{s}^2}{\omega_\text{p}^2A_\text{p}^2},
\end{align}
where $G(A_\text{s})$ is the power gain as a function of the input signal amplitude $A_\text{s}$, $A_\text{p}$ is the amplitude of the pump, and $\omega_\text{s}$ and $\omega_\text{p}$ refer to the signal and pump frequencies, respectively. Here, $P_\tin$ is the power of the input signal and $P_\text{pump}$ is the power of the pump wave. We note that $P_\text{out}$ consists only of the power being converted as output at the signal frequency, while ignoring the power at the idler frequency. Therefore, for nondegenerate amplifiers, but not degenerate amplifiers, approximately half of the output power, which comes out at the idler frequency, is not accounted for in the PAE. We choose this definition because we expect that only the signal frequency will be used downstream. We also note that in this work, we will focus on amplifiers with a target gain of $G_\text{t}=20~\dB$. Finally, we note that for convenience of the reader, we list commonly used symbols in Tab.~\ref{tab:symbol}.

We define the PAE of an amplifier, $\eta_\text{PAE}$, to be the maximum value of $\eta_\text{PAE}(A_\text{s})$ in the range of $A_\text{s}$ below saturation amplitude, $A_{\text{sat}}$. A higher PAE is desirable in JPAs used in large scale quantum computing applications to minimize heat load on the fridge.
However, modern JPA designs tend to have pump efficiencies $(P_{\text{out}}/P_{\text{pump}})$ of less than  $\sim$0.1\% \cite{sivak2019}, indicating a huge gap between desired and actual performance levels. 

\begin{figure}
    \centering
    \includegraphics[width = 0.9\linewidth]{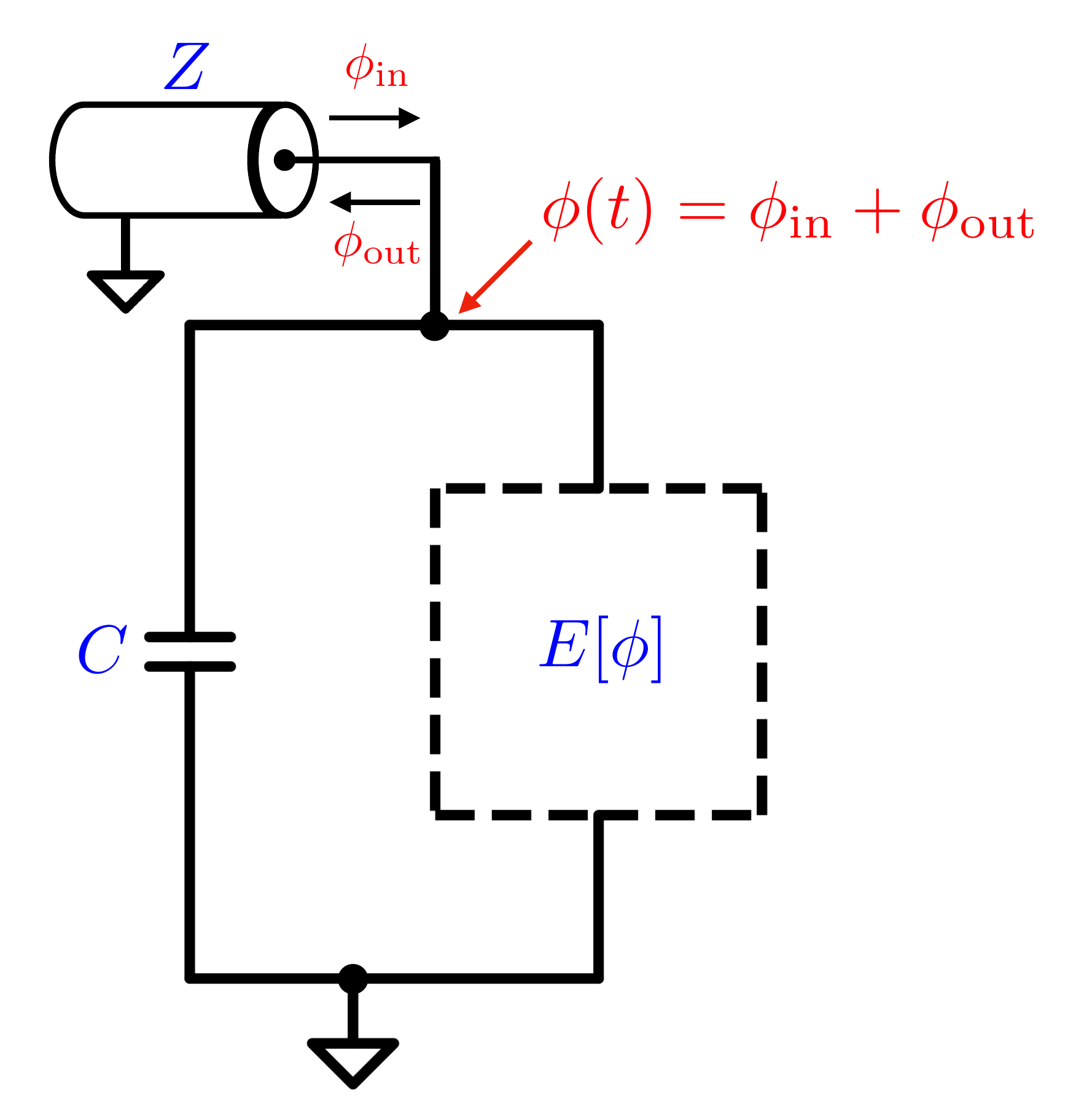}
    \caption{\label{fig:HCircuit}Generic amplifier circuit coupled to a transmission line with impedance $Z$. The circuit is composed of a capacitor $C$ and an inductive block with energy $E[\phi]$. The amplifier phase $\phi(t)$ is coupled to the transmission line via the input/output relation as indicated.}
\end{figure}

In this paper, we will investigate the ultimate limit on the PAE of a class of parametric amplifiers illustrated in Fig.~\ref{fig:HCircuit}. We will show that amplifiers within this class can have a PAE orders of magnitude higher than typical JPA designs. Specifically, we consider amplifier circuits of the general form shown in Fig.~\ref{fig:HCircuit} that consist of a capacitor (denoted by $C$) in parallel with a nonlinear inductive block with energy $E[\phi]$ \footnote{Note that we do not consider a coupling capacitor to the transmission line, as is commonly used in experiment. This is because coupling capacitors reduce effective PAE and we are interested in the PAE of the amplifier after the plane of the coupling capacitor.}.
We first use this framework to explore the ultimate limit on PAE by considering arbitrary functions $E[\phi]$. Next, we explore buildable circuits where $E[\phi]$ represents the energy of a collection of inductors and Josephson junctions. We also introduce a rule for designing amplifier circuits -- namely that the current-phase relation of the inductive block should be monotonic. Finally, we explore improving PAE by using amplifier chains. In the next three paragraphs, we summarize the results of these three directions.

To explore the ultimate limit on PAE, we begin by writing the energy of the inductive block of the amplifier as a generic polynomial
\begin{align}
E[\phi]=\frac{1}{2L_\text{eff}}\phi^2 +g_3\phi^3+g_4\phi^4+\cdots,
\end{align}
where $L_\text{eff}$ is the effective inductance of the inductive block.
We tune the $g_i$ coefficients in this polynomial amplifier in order to maximize saturation power and, consequently, PAE. Searching through the space of polynomial amplifiers sets a bound on what level of efficiency is achievable given a certain order of nonlinearity, i.e., we can increase the PAE by adding further terms to $E[\phi]$. We have explored up to tenth order amplifiers. In Subsec.~\ref{sec:PolyAmpResult}, we find that the PAE tends to saturate at around sixth order, where the maximum achievable PAE is at least 63\% for degenerate amplifiers with $G_\text{t}=20~\dB$.

Next, we investigate whether it is possible to achieve the same PAE from buildable amplifiers with inductive blocks composed of inductors and Josephson junctions. We replace the generic $E[\phi]$ by the energy of a circuit composed of these circuit elements. We choose circuits which have a comparable number of tuning parameters as in the polynomial case and adjust these parameters to optimize saturation power. In Subsec.~\ref{sec:JPACircuits}, we show that we can reach a maximum PAE of 42.6\% for degenerate amplifiers with $G_\text{t}=20~\dB$.

Finally, we explore increasing PAE by chaining together amplifiers. The PAE of an amplifier usually decreases as the gain increases. Therefore, one might expect to obtain an amplifier with a higher PAE by connecting several smaller gain amplifiers together. In Subsec.~\ref{sec:chainAmp},
we find that we can increase the PAE at $20$~dB from around $60\%$ to more than $98\%$ by chaining $\sim 15$ amplifiers (see in Fig.~\ref{fig:chainAmpsEtavsN}).

\section{Amplifier Model and Numerics}

\begin{figure}
    \centering
    \includegraphics[width = \linewidth]{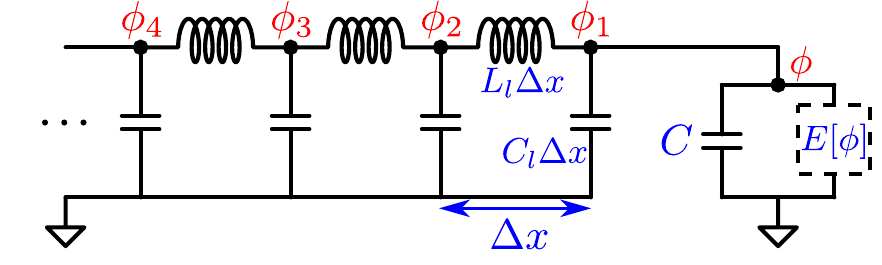}
    \caption{Schematic of the parametric amplifier with an explicit model for the transmission line, which we use to illustrate the origin of the input/output relations. Here, $\phi, \phi_1, \phi_2,\cdots$ are node fluxes in the transmission line; $C_l,L_l$ are capacitance and inductance per unit length, respectively; $\Delta x$ is the cell length. }
    \label{fig:PolyAmp}
\end{figure}

In this section, we lay out the model for a class of parametric amplifiers composed of a capacitor shunted by a nonlinear inductive block (shown in Fig.~\ref{fig:PolyAmp}), derive the classical input-output relation, and obtain the equation of motion (EOM). Our calculations mainly follow Ref.~\onlinecite{vool2017introduction}.

\subsection{Model}

Here the amplifier is made of a capacitor and a nonlinear inductive block whose energy is $E[\phi]$. The Lagrangians of the amplifier, $\mathcal{L}_{\text{s}}$, and the transmission line, $\mathcal{L}_{\text{tl}}$, are
\begin{align}
    \mathcal{L}_{\text{s}} &= \frac{C\dot{\phi}^2}{2} - E[\phi], \\
    \mathcal{L}_{\text{tl}} &= \sum_{i=1}\left(\frac{C_l \, \Delta x \, \dot{\phi}_{i+1}^2}{2} - \frac{({\phi_{i+1}}-{\phi_{i}})^2}{2L_l \, \Delta x}\right).
\end{align}
To couple the transmission line with the amplifier, we set $\phi_1 = \phi$.

\subsection{Equations of motion and input-output relations}

We first consider the transmission line except for the last node $\phi_1$. For $i> 1$, the equation of motion for $\phi_i$ is given by
\begin{align}
\begin{aligned}
    C_l \, \Delta x \,  \ddot{\phi}_i + \frac{{\phi_i}-{\phi_{i-1}}}{L_l \, \Delta x}-\frac{{\phi_{i+1}}-{\phi_{i}}}{L_l \, \Delta x} = 0,
\end{aligned}
\end{align}
which in the continuum limit, $\Delta x \to 0$, becomes the wave equation,
\begin{align}
     \ddot{\phi} + \frac{\phi''}{C_lL_l} = 0,
\end{align}
where $'$ denotes the spatial derivative.
We define the incoming and outgoing waves $\phi_{\tin}$ and $\phi_{\out}$ such that
\begin{align}
    &\left(\frac{\partial}{\partial t}+\frac{1}{\sqrt{C_lL_l}}\frac{\partial}{\partial x} \right) \phi_{\tin} = 0, \\
    & \left(\frac{\partial}{\partial t}-\frac{1}{\sqrt{C_lL_l}}\frac{\partial}{\partial x} \right) \phi_{\out} = 0.
\end{align}
Then we have
\begin{align}
    &\phi(x,t) = \phi_{\tin}(x,t)+\phi_{\out}(x,t).
\end{align}

Next, we consider the boundary node flux $\phi_1 = \phi$, which gives the input-output relation
\begin{align}
    \phi(t) = \phi(0,t) = \phi_{\out}(0,t)+\phi_{\tin}(0,t). \label{eq:input-output}
\end{align}
The EOM of the boundary node flux is given by
\begin{align}
    {C\ddot{\phi}}  +  J[\phi] = \frac{\phi'|_{x=0}}{L_l} = \left. \frac{\dot{\phi}_{\tin}-\dot{\phi}_{\out}}{Z}\right|_{x=0},
\end{align}
where $Z = \sqrt{L_l/C_l}$ is the characteristic impedance of the transmission line and  $J[\phi] = \dd E/ \dd \phi$ is the current through the inductive block as a function of the phase across the block, $\phi$.
Together with the input-output relation~\eqref{eq:input-output}, the evolution of the nonlinear system becomes
\begin{align}
    \ddot{\phi} + K\dot{\phi} +\frac{J[\phi]}{C}=2K\dot{\phi}_{\tin}|_{x=0},\label{eq:NonlinearEOM}
\end{align}
where $K = \frac{1}{CZ}$ is the damping rate of the amplifier from its coupling to the transmission line. 

\subsubsection{Polynomial amplifiers}

A polynomial amplifier of order $n$ is defined by a polynomial energy,
\begin{align}
     E[\phi] = \frac{1}{2L_\text{eff}} \phi^2 + g_3\phi^3+\cdots + g_n\phi^n. \label{eq:ployEng}
\end{align}
To ensure stability of the amplifier, we require $n = 2m$ to be a even number and $g_n>0$, to guarantee the existence of a global energy minimum. Ideally, we want $\phi = 0$ to be the unique minimum point of the energy function, but this is often impractical. Instead, we demand that in the vicinity of $\phi=0$, $E[\phi]$ is a monotonically increasing function of $\phi$ as we move away from the origin in either direction.

The corresponding current-phase relation $J[\phi]$ is a polynomial of order $n-1$,
\begin{align}
    J[\phi] = \frac{1}{L_\text{eff}} \phi + 3g_3\phi^2 +\cdots + ng_n \phi^{n-1}. \label{eq:CPR}
\end{align}
The equation of motion for this amplifier is given by 
\begin{align}
\begin{aligned}
    \ddot{\phi} + K\dot{\phi}+\omega_0^2{\phi} +\frac{1}{C}(3g_3\phi^2+&\cdots+ng_{n}\phi^{n-1}) \\
    &={2K\dot{\phi}_{\tin} }|_{x=0},\label{eq:EOM}
\end{aligned}
\end{align}
where $\omega^2_0 = \frac{1}{CL_\text{eff}}$ is the natural frequency of the amplifier. We note that in  this polynomial amplifier, the third order ($g_3$) term is primarily responsible for amplification. However, the third order term dynamically generates higher order nonlinear terms which limit the amplifier's performance at high signal power~\cite{liu2017josephson,liu2020JRM}. We aim to cancel the dynamically generated terms with higher order static terms in Eq.~\eqref{eq:EOM} in order to increase the saturation power and hence the PAE of the amplifier at fixed pump amplitude.

\subsubsection{Josephson parametric amplifiers} \label{sec:JPAdesign}

In this subsection, we will discuss how to construct the inductive block $E[\phi]$ from linear inductors and Josephson junctions. We start from the Lagrangian description of inductors and Josephson junctions. The contributions of these elements to the energy $E[\phi]$ are
\begin{align}
E_L&=\frac{\phi_0^2}{2L}(\varphi_1-\varphi_2)^2,\\
E_J&=-\phi_0 i_\text{c} \cos(\varphi_1-\varphi_2),
\end{align}
where $E_L,E_J$ are the contributions of inductors and Josephson junctions, respectively. Here, $L$ is inductance, $i_\text{c}$ is Josephson critical current, and $\phi_0$ is the reduced magnetic flux quantum. For JPAs we use dimensionless fluxes $\varphi_i=\phi_i/\phi_0$. In addition, we consider both current and flux biases in these circuits.

We represent the inductive block $E[\varphi]$ by a network of internal inductive components which make up our circuit design. Specifically, the circuit is composed of a set of internal nodes $\varphi_i$ that are connected by the inductive components. In Ref.~\onlinecite{Rymarz2023}, it was shown that stray capacitances could have important effects. We discuss the effect of capacitive terms within the inductive block in Subsec.~\ref{sec:modeCouple}.

For each node $\varphi_i$, we write that $\frac{1}{\phi_0} \frac{\partial \mathcal{L}}{\partial \varphi_i}=J_i=J_{\text{ext},i}$ where $J_{\text{ext},i}$ is the current bias at node $\varphi_i$. Typically, nodes are not current biased and thus have $J_i=0$. From here we proceed as in Ref.~\onlinecite{liu2020JRM}.

We also bias the circuit by applying external magnetic flux through closed loops in the circuit. Without loss of generality, we model magnetic flux bias by describing it as a phase offset across one of the Josephson junctions in the loop. In particular, a Josephson junction in a loop with an applied external flux of $\Phi_{\text{ext}}$ will have a contribution of $\frac{\phi_0^2}{2L}(\varphi_1-\varphi_2+\varphi_{\text{ext}})^2$ to the Lagrangian, where $\varphi_{\text{ext}}=\Phi_{\text{ext}}/\phi_0$.

\begin{figure}
    \centering
    \includegraphics[width = 0.9\linewidth]{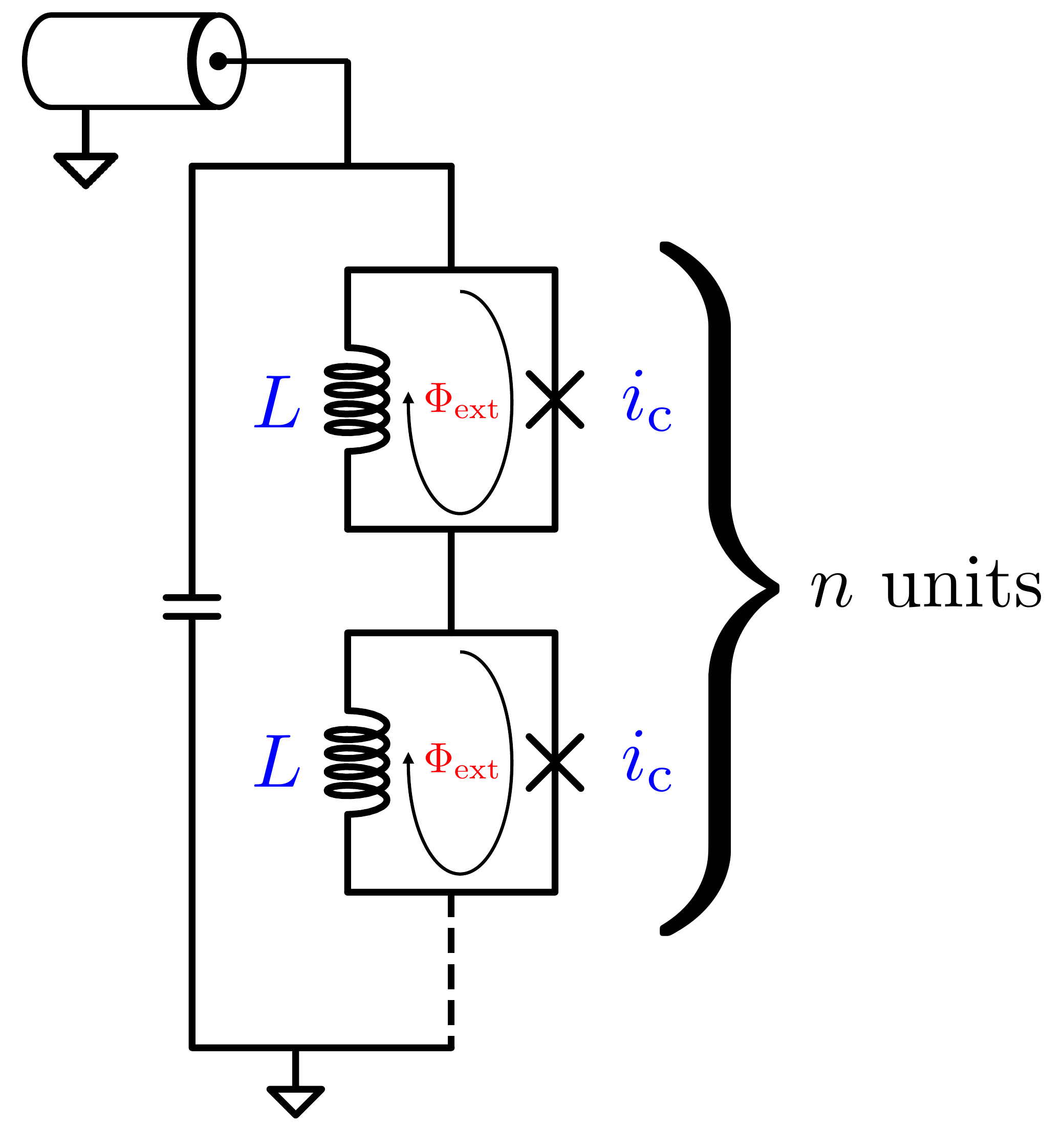}
    \caption{\label{fig:RFSquidAmp}Amplifier circuit with an inductive block composed of a chain of $n$-many RF-SQUIDs.}
\end{figure}

To make these considerations concrete, we examine the case of a JPA with an inductive block composed of $n\text{-many}$ radio frequency superconducting quantum interference devices (RF-SQUIDs)~\cite{White2023,Kaufman2023,Kaufman2024} as shown in Fig.~\ref{fig:RFSquidAmp}. In this case, we will for now describe only a single RF-SQUID making the assumption that the flux $\varphi$ across the inductive block is divided equally among the $n$-many identical SQUIDs, and therefore each SQUID experiences a flux of $\frac{\varphi}{n}$. In Subsec.~\ref{sec:modeCouple}, we investigate this assumption and show that it holds in the case that the RF-SQUIDs have a monotonic current-phase relation. The resulting equation of motion is
\begin{align}
\begin{aligned}
    \ddot{\varphi}+K\dot{\varphi}+\frac{\varphi}{nLC}+\frac{i_\text{c}}{\phi_0 C} \sin\left( \frac{\varphi}{n}+\varphi_{\text{ext}}\right)=2K\dot{\varphi_{\tin}},\label{eq:rfsquid}
\end{aligned}
\end{align}
where $L$, $i_\text{c}$, and $\varphi_{\text{ext}}$ are the linear inductance, Josephson critical current, and dimensionless external flux through each RF-SQUID, respectively.
In general, we can construct the current-phase relation $J[\varphi]$ for the designed amplifier circuit and apply it to the EOM specified in Eq.~\eqref{eq:NonlinearEOM}.

\subsection{Numerical solutions of EOM and optimization algorithm for PAE}\label{sec:numerics}

In this section, we provide the optimization algorithm that we use to maximize the PAE in the presence of the following incoming wave 
\begin{align}
    \phi_{\tin}|_{x=0} = A_\sig \sin(\omega_\sig t) + A_\pump \sin(\omega_\pump t+\delta),
\end{align}
where $\omega_\sig, \omega_\pump$ are the signal and pump frequencies, $A_\sig, A_\pump$ are the signal and pump amplitudes, and $\delta$ is the phase difference between the signal and pump wave. Because the differential equation~\eqref{eq:EOM} is nonlinear, inhomogeneous and non-autonomous, obtaining analytical solutions is not possible in general. Thus, we first introduce two numerical methods, the direct time integration (DTI) and the harmonic balance (HB) method, which we use to solve this differential equation.

The DTI method is based on numerical integration techniques that approximate the differential equation by using finite differences. The DTI method uses function values and finite differences at current and previous time intervals to find the value at the next interval. In our calculation, we use the Python method \textit{scipy.integrate.odeint}~\cite{SciPy} to solve the nonlinear differential equation.
Since we want to obtain a steady-state solution, we choose our integration time to be at least 2000 periods of the signal wave. Then, to remove the initial transient, we only analyze the data from the last quarter of the integration time.
After obtaining the solution in the time domain, we apply the discrete Fourier transform (DFT) 
to isolate the outgoing signal wave from the rest of the solution. This requires that the last quarter of the integration time is a multiple of $T_{\min} = \text{lcm}(\frac{2\pi}{\omega_\sig},\frac{2\pi}{\omega_\pump})$, where $\text{lcm}$ is a function giving the least common multiple of its inputs, which dictates our choice of $\omega_\pump/\omega_\sig$. Finally, the PAE is determined by calculating both the saturation power gain and the amplitude of the signal wave at that gain.

The HB method exploits the periodicity of the external driving force and solves the differential equation in the frequency domain. If the incoming wave $\phi_{\tin}$ is periodic with a period of $T$, then the solution to Eq.\eqref{eq:EOM} is also periodic with the same period $T$.
Thus, we can propose that the solution to Eq.~\eqref{eq:EOM} takes the following form (in the degenerate case):
\begin{align}
    \phi(t) = a_0 + a_{1}e^{i\omega_\sig t} + \cdots + a_{n}e^{in\omega_\sig t} + h.c..
\end{align}
Then, the solution can be obtained by matching coefficients on both sides of Eq.~\eqref{eq:EOM}. To ensure that we have chosen a sufficiently large $n$, we also compare the HB method with the DTI method. 
For degenerate amplifiers, the HB method is typically much faster than the DTI method. On the other hand, for nondegenerate amplifiers, we need to consider many more harmonics in the HB method. In this case, the DTI method is faster, and is the one we use.
A more comprehensive analysis of these methods is provided in Appendix~\ref{appx:HBmethod}.

Before we describe the optimization algorithm, we normalize the EOM by rescaling time and flux. By doing so, we reduce the number of parameters in the differential equation, making it easier to work with. The initial differential equation is 
\begin{align}
    \ddot{\phi} + K\dot{\phi}+\omega_0^2{\phi} +\frac{1}{C}(3g_3\phi^2+\cdots+ng_{n}\phi^{n-1}) ={2K\dot{\phi}_{\tin} },
\end{align}
with the incoming wave $\phi_{\tin} = A_\sig \sin(\omega_\sig t) + A_\pump \sin(\omega_\pump t+\delta)$. First, we rescale time $t \to 2t/\omega_\text{p} $, i.e., we work in units where $\omega_\text{p} = 2$. Next, we scale flux so that $A_\pump$ is normalized to be $0.5$. In this case, the pump power is $A_\pump^2\omega_\pump^2 = 1$ and the PAE is $(G-1)\omega_\sig^2A_\sig^2$. Finally, we rescale coefficients $g_i$ to absorb $1/C$. In the following, we will use the same letter to denote the normalized polynomial coefficients. After rescaling, our differential equation becomes
\begin{align}
    \ddot{\phi} + K{ \dot{\phi}}+ \omega_0^2{\phi} +3g_3\phi^2+\cdots+ng_{n}\phi^{n-1}= 2K\dot{\phi}_{\tin} ,
    \label{eq:polyEOM}
\end{align}
with the incoming wave $\phi_{\tin} = A_\sig \sin(\omega_\sig t) + \frac{1}{2} \sin(2t+\delta)$. 

\begin{figure}
    \centering
    \includegraphics[width = 0.9\linewidth]{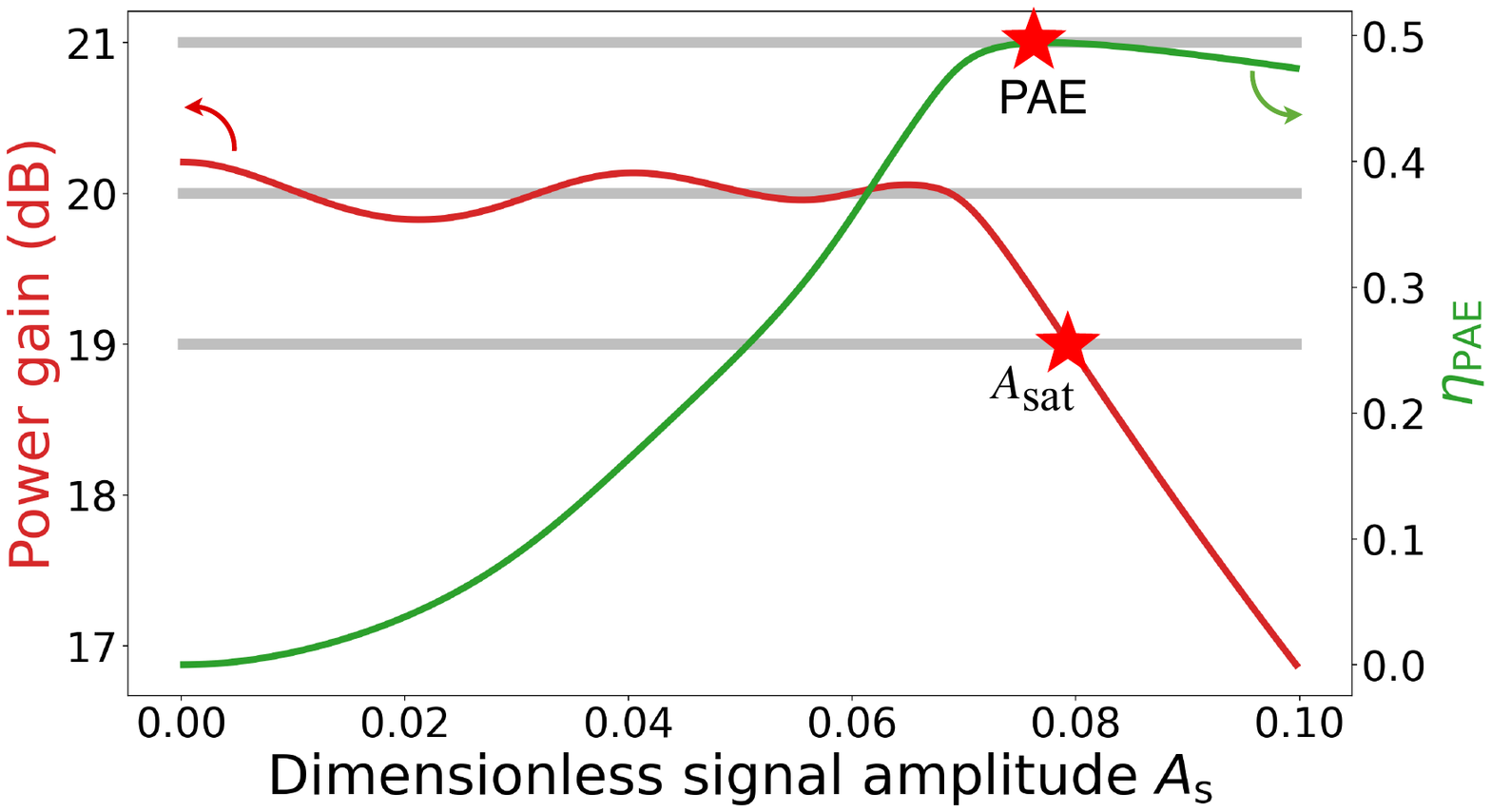}
    \caption{Power gain and PAE versus dimensionless signal amplitude $A_{\text{s}}$ of a 6-th order 20~dB degenerate amplifier. The parameters of this amplifier can be found in Tab.~\ref{tab:optimization}. The saturation amplitude $A_{\text{sat}}$ is the signal amplitude where the gain curve intersects the $G_\text{t}\pm 1$~dB lines. The PAE of an amplifier is defined to be the maximum PAE below the saturation power, which typically appears at or around $A_{\text{sat}}$.}
    \label{fig:PAE}
\end{figure}

\begin{figure}[b]
    \centering
    \includegraphics[width = \linewidth]{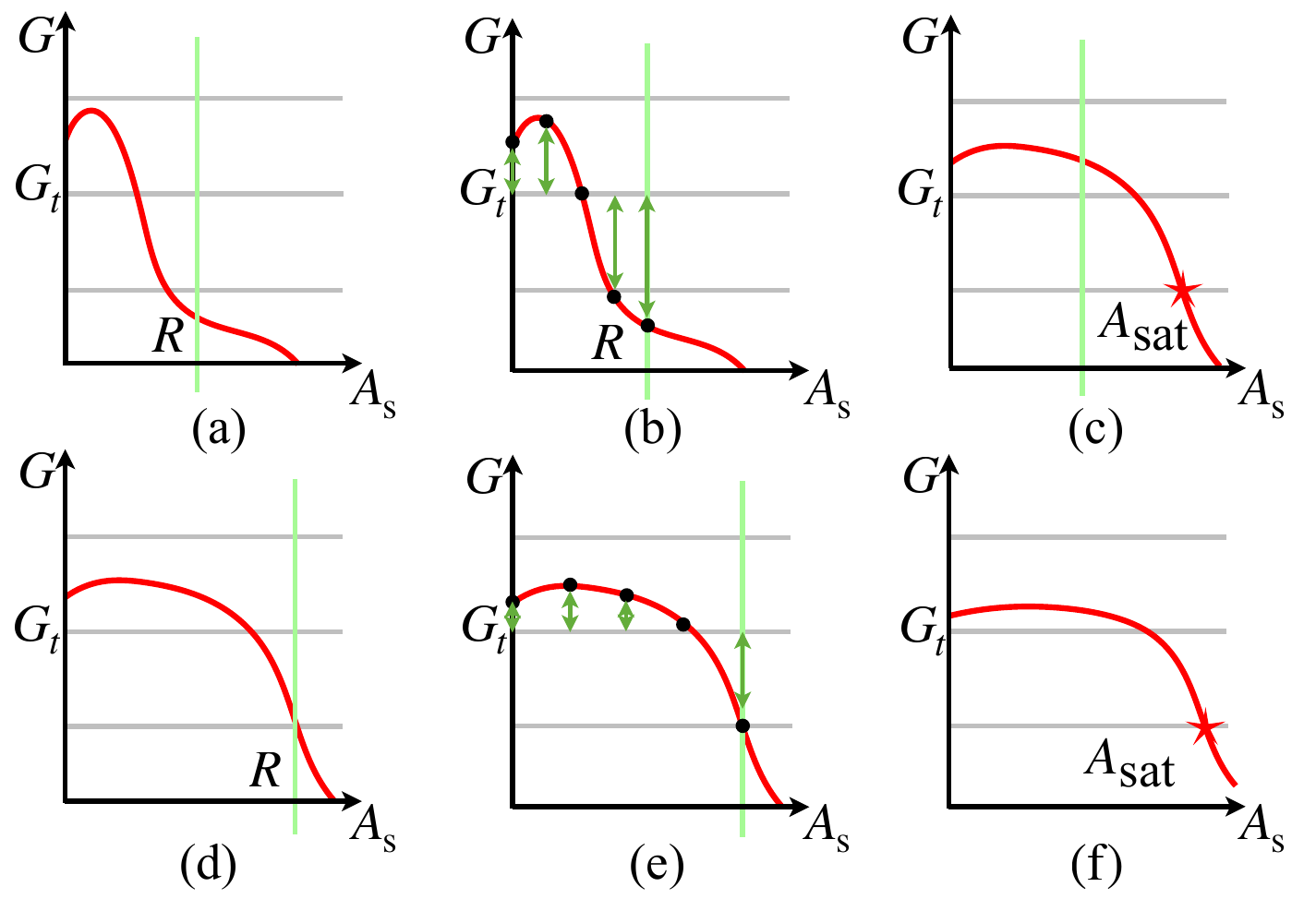}
    \caption{Schematic of optimization algorithm steps. (a)~Start from an initial optimization range $[0, R]$ and a set of initial variables $\{\omega_0^{(0)}, K^{(0)}, g^{(0)}_3, g^{(0)}_4, \cdots\}$. (b)~Calculate the cost function corresponding to the green arrows. (c)~Minimize the cost function. (d)-(e)~Update the optimization range $[0, R]$ and re-optimize the amplifier's coefficients until the optimization process stops. (f)~Record the optimal values of the amplifier parameters $\{\omega_0, K, g_3, g_4, \cdots\}$.}
    \label{fig:update}
\end{figure}

While we wish to maximize the PAE of the amplifier directly, this is difficult, and therefore we use a proxy. Specifically, we minimize the difference between the amplifier gain and $G_\text{t}$ over the largest possible range of $A_\sig$. We take the highest PAE below the saturation power of the amplifier as the amplifier's PAE. This maximum PAE typically occurs at or near the saturation power of the amplifier (see Fig.~\ref{fig:PAE}).
The optimization variables we use are $\omega_0, K, g_3, g_4, \cdots$. We use $\omega_0$ as an optimization variable since it is possible that the amplifier should have a natural frequency which is different than the target signal frequency $\omega_s$~\cite{liu2017josephson}. In the remainder of this section, we will present the algorithm step by step with necessary explanations. A pictorial demonstration of the algorithm is also provided in Fig.~\ref{fig:update}.

\textbf{Step 1.} Choose an initial range $[0,R]$ for $A_\text{s}$, and a set of initial variables $\{\omega_0^{(0)},K^{(0)}$, $g_3^{(0)}$, $g_4^{(0)},\cdots\}$. This set of variables generates an initial gain curve as shown in Fig.~\ref{fig:update}(a).   

\textbf{Step 2.} Divide the region $[0,R]$ into $N$ sub-intervals and define the cost function $f$ as the sum of the squared distance (illustrated in Fig.~\ref{fig:update}(b)) plus the penalty that ensures that $E[\phi]$ is a monotonically increasing function of $\phi$ as one moves away from $\phi=0$ in either direction. For finer optimization, we use an alternative cost functions. For more information on this and the applied penalty, see Appendix~\ref{appx:enforce}. We have checked that the range over which we enforce the condition on $E[\phi]$ is larger than the largest fluctuations of $\phi(t)$ that we observe. 

\begin{align}
\begin{aligned}
    &f\left(\omega_0^{(0)},K^{(0)},g_3^{0},\cdots\right) \defeq  \\
    & \sum_{m=1}^{N}\left(G\left(A_\sig = \frac{mR}{N} ,\omega_0^{(0)},K^{(0)},g_3^{(0)},\cdots\right) - G_\text{t} \right)^2 + \text{penalty} 
\end{aligned}
\label{eq:CostFunction}
\end{align}

\textbf{Step 3.} Minimize the cost function $f$ using the gradient descent algorithm. By doing so, we will get a new collection of coefficients $\{\omega_0^{(1)},K^{(1)}$, $g_3^{(1)}$, $\cdots\}$.

\textbf{Step 4.} Calculate the saturation point for the new curve, which is shown in Fig.~\ref{fig:update}(d), and reset the new range to be $[0, A_{\text{sat}}]$ if $A_{\text{sat}}$ exceeds the previous range $R$; otherwise, stop the optimization.

\textbf{Step 5.} Redefine the cost function $f$ within the new range as depicted in Fig.~\ref{fig:update}(e), and repeat steps 3 to 5 until the optimization stops.

\textbf{Step 6.} Record the optimized values of the amplifier parameters $\omega_0,K, g_3, \cdots$.

The solution and optimization of amplifiers described by circuits is similar to the description above, and we provide details of this procedure in Appendix~\ref{appx:JPA}.

\section{Results}

\subsection{Polynomial amplifiers}\label{sec:PolyAmpResult}

In this subsection, we apply our optimization algorithm to the polynomial amplifier to find a lower bound on the maximum PAE for the class of amplifiers that we consider in Fig.~\ref{fig:PolyAmp}. The goal of this subsection is to set a benchmark for amplifiers that can be expressed as circuits, which we study in the next subsection.
We first focus on degenerate amplifiers where $2\omega_\text{s} = \omega_\text{p}$. We fix the relative phase $\delta$ to be $1.5\pi$. This choice tends to maximize the gain in the small signal regime.
We will start by establishing how the PAE depends on the order of the polynomial by optimizing a set of amplifiers with different orders to the target gain of 20~dB. We will choose the polynomial order that balances the algorithm's run time and amplifier performance. 
Next, we will fix the polynomial order and vary the target gain to determine the relation between target gain and PAE. Finally, we will switch our focus to nondegenerate amplifiers and determine the dependence of PAE on target gain.

\begin{figure}
    \centering
    \includegraphics[width = 0.9\linewidth]{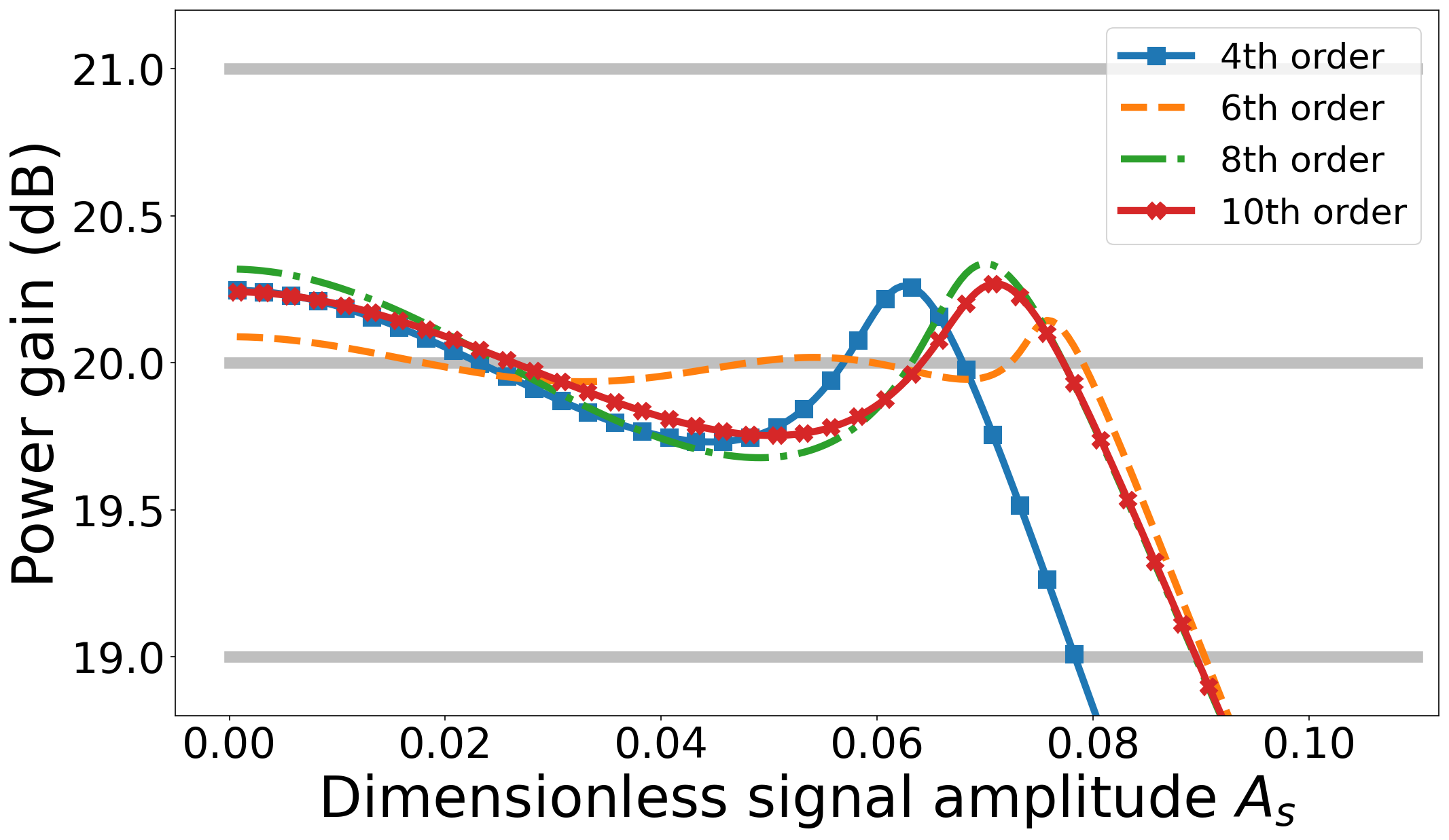}
    \caption{Power gain versus dimensionless signal amplitude curves for optimized 20~dB polynomial amplifiers with different orders of nonlinearity. Parameters of these amplifiers are listed in Tab.~\ref{tab:optimization}.}
    \label{fig:Gvsorder}
\end{figure}

We start by optimizing the PAE of $20~\dB$ degenerate polynomial amplifiers with order varying from 4 to 10.
The outcomes of this process are illustrated in Fig.~\ref{fig:Gvsorder}. To avoid local optima, we repeat our optimization algorithm with different starting points multiple times for each target gain and keep the highest PAE. We observe that as we increase the polynomial order from 4 to 10, the saturation power $A_{\text{sat}}$ of these amplifiers improves significantly, 
rising from about 0.08 to around 0.09, corresponding to a rise in PAE from approximately 50\% to around 60\%. We note that the PAE saturates at around the sixth order (see in Tab.~\ref{tab:etavsOrder}). So, we focus on the performance of the 6th-order polynomial amplifier in this section.

\begin{table}[]
    \centering
    \begin{ruledtabular}
    \begin{tabular}{ccc}
        polynomial order & dimensionless $A_{\text{sat}}$ & PAE \\
        \hline
        4&0.0783 & 48.0\%\\
         6&0.0845 & 63.3\%\\
         8&0.0894&62.6\%\\
         10&0.0895& 62.7\%
    \end{tabular}
    \end{ruledtabular}
\caption{\label{tab:etavsOrder} Saturation amplitude $A_{\text{sat}}$ and PAE of optimized 20~dB polynomial amplifiers with different orders of nonlinearity. Parameters of the amplifiers can be found in Tab.~\ref{tab:optimization}.}
\end{table}

\begin{figure}
    \centering
    \includegraphics[width = 0.9\linewidth]{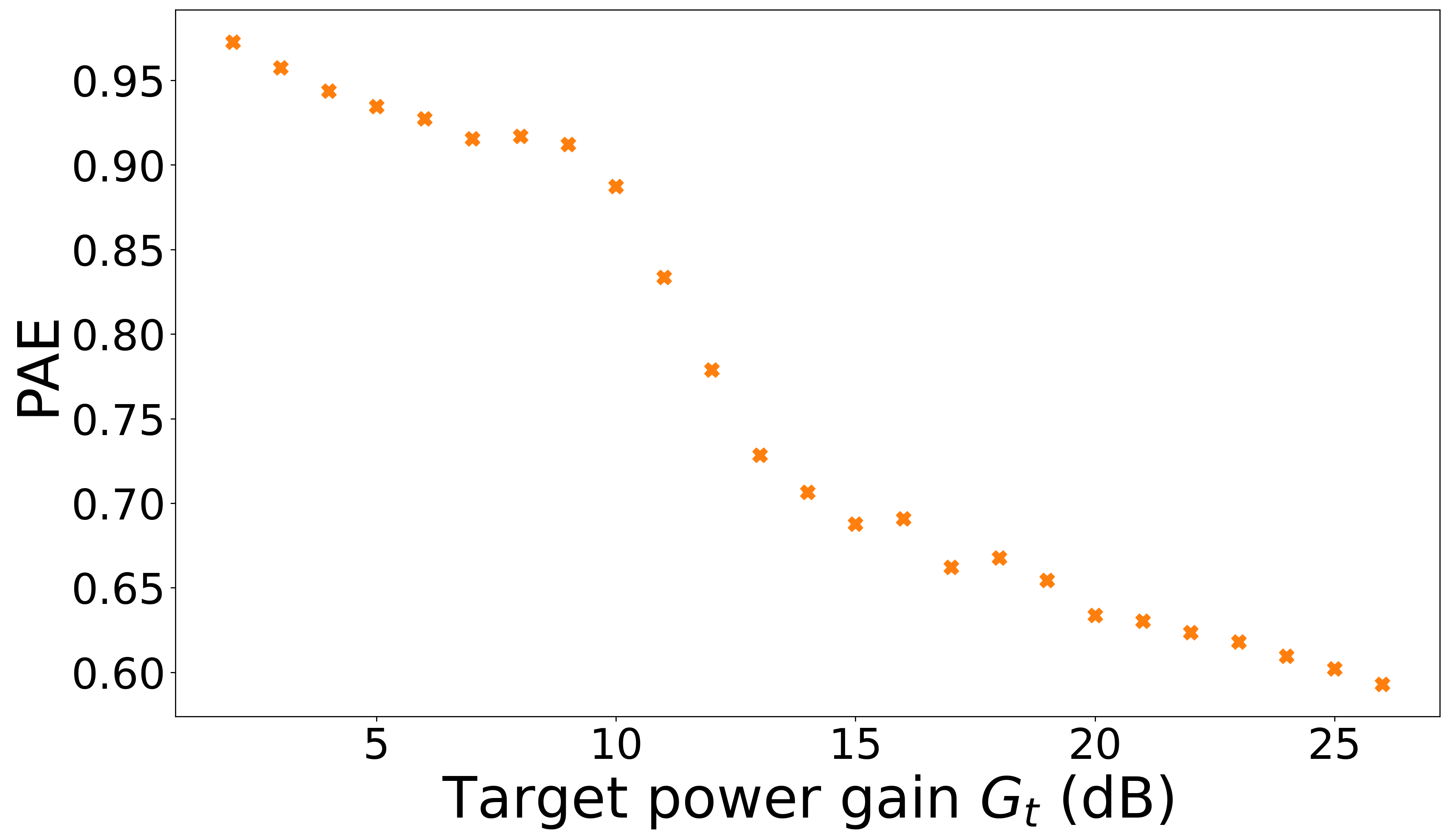}
    \caption{PAE versus target gain of $6$-th order polynomial degenerate amplifiers, showing that PAE generally decreases as amplifier gain increases.  The kink around $G_t = 10~\text{dB}$ results from the penalty imposed in the cost function~\eqref{eq:CostFunction}. For small target gains, this penalty is zero and does not affect the optimization process. As the target gain increases, the penalty restricts the design from achieving high PAE in order to prevent instability of the amplifier.}
    \label{fig:etavsGt}
\end{figure}

\begin{figure}
    \centering
    \includegraphics[width = 0.9\linewidth]{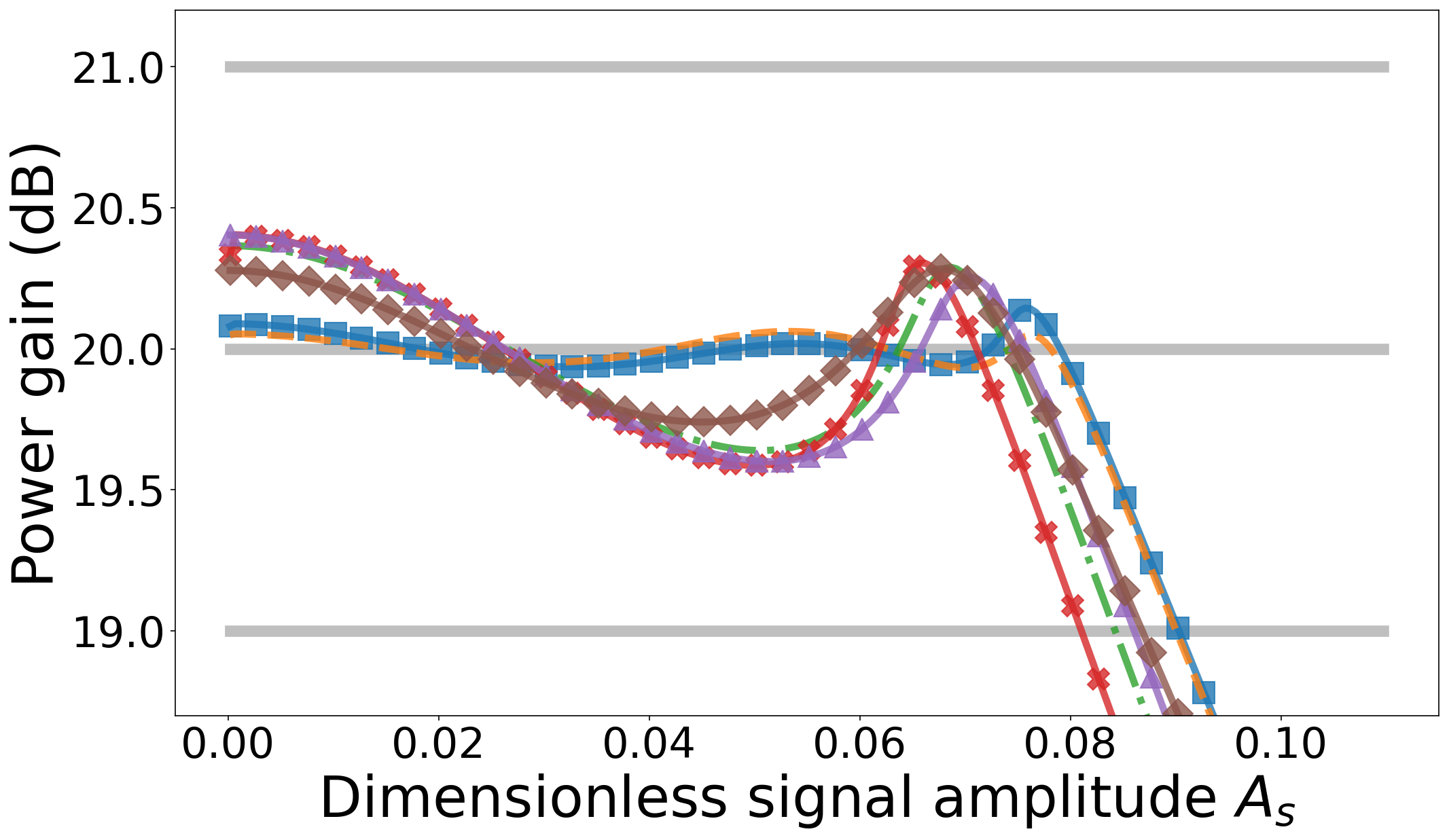}
    \caption{Importance of initial points to optimization results. Here, we show gain versus signal amplitude curves of 20~dB $6$-th order polynomial amplifiers, where different curves represent amplifiers that were found by optimizing from six different starting points.
    }
    \label{fig:20dBamplifiers}
\end{figure}

Next, we sweep the target gain of the 6th-order polynomial degenerate amplifiers from 2~dB to 26~dB. 
As depicted in Fig.~\ref{fig:etavsGt}, there is a gradual decline in the PAE, dropping from nearly 100\% at 2~dB to about 60\% at 26~dB. 
As before, we use multiple starting points for the optimization algorithm. In Fig.~\ref{fig:etavsGt} we plot the resulting PAEs.
At lower target gains, different starting points generally yield similar PAEs.
However, in the high gain regime, we observe that the algorithm's output begins to fluctuate, with some amplifiers tending to be trapped in local optima, as shown in Fig.~\ref{fig:20dBamplifiers}. 

We used the optimal points of the degenerate amplifiers as starting points for the optimization of the nondegenerate amplifiers. The reason for this choice, as opposed to using a large number of random starting points, was that the large computational complexity of characterizing nondegenerate amplifiers made trying a large number of starting points computationally too expensive. We remind the reader that characterizing nondegenerate amplifiers is computationally much more expensive than characterizing degenerate amplifiers because the former requires the use of the direct time integration method while the latter can be much more efficiently handled by the harmonic balance method. Even within the context of the direct time integration method, nondegenerate amplifiers require longer integration time since $T_{\min}$ is large. This is due to the slight offset of the signal frequency from half the pump frequency, which increases $\text{lcm}(\frac{2\pi}{\omega_\sig},\frac{2\pi}{\omega_\pump})$.

\begin{figure}
    \centering
    \includegraphics[width = 0.9\linewidth]{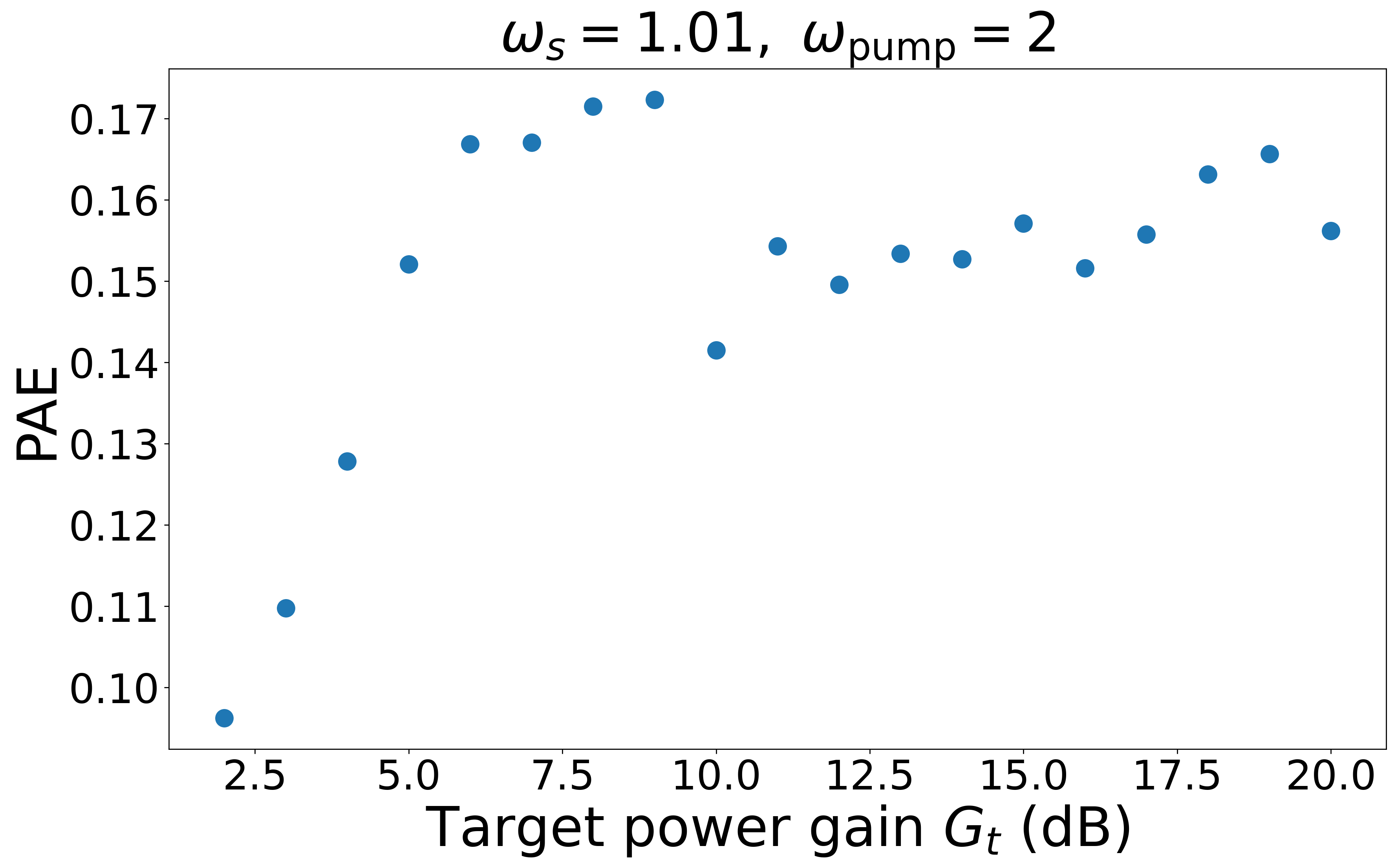}
    \caption{Pump efficiency as a function of target gain for $6$-th order polynomial nondegenerate amplifiers. The PAE increases from about 10\% at 2~dB to around 15\% at 20~dB. }
    \label{fig:etavsGtND}
\end{figure}

For the case of degenerate amplifiers, we previously observed a steady decline of the PAE with increasing target gain (see Fig.~\ref{fig:etavsGt}). Conversely, for nondegenerate amplifiers, the PAE increases from about 10\% at 2~dB to around 15\% at 20~dB where it saturates (see Fig.~\ref{fig:etavsGtND}). This counterintuitive behavior can be elucidated as follows. Consider a nondegenerate amplifier in the degenerate limit of $\delta \omega = (\omega_\sig - \omega_\mathrm{i})/2 \to 0$. In this case, the output is a single wave whose amplitude is a combination of the signal and idler components, $A = A_\sig + A_\mathrm{i}$. Since the power is proportional to the square of the amplitude, the signal wave power gain for the nondegenerate amplifier is four times ($\sim$ 6~dB) smaller than that of the corresponding degenerate amplifier\footnote{This factor of 4 is a consequence of our definition of PAE, rather than a fundamental disadvantage of nondegenerate amplification.}. 

For a degenerate amplifier at signal amplitude $A_s$ and frequency $\omega_\sig$ with a gain $G$, the PAE is expressed as: $\eta(G) = (G-1)\omega_{\mathrm{s}}^2A_\mathrm{s}^2$. The PAE for the corresponding nondegenerate amplifier with signal frequency $\omega_\sig + \delta \omega$ and gain $G/4$ is:
\begin{align}
    \eta'\left(\frac{G}{4}\right) = \left(\frac{G}{4}-1\right)(\omega_\mathrm{s} + \delta \omega)^2A_\text{s}^2 \approx  \frac{G-4}{4G-4}\eta(G).
\end{align}
This leads to $\eta' \to 0$ as $G/4 \to 1$. Additionally, if $\eta(G)$ is a constant, $\eta'$ increases monotonically, approaching $\eta(G)/4$.  Thus, we can also get the upper bound for the nondegenerate PAE, $\eta' \leq 25\%$, where the equality holds when $G \to \infty$ and $\eta(G) = 1$.
We remind the reader that our definition of PAE only accounts for the power leaving the amplifier at the signal frequency and not the idler frequency.

\subsection{Coupling of low- and high-frequency modes in JPA circuits}\label{sec:modeCouple}

Our amplifier circuits are generally composed of $n$-many identical repeating elements which equally divide the phase across the amplifier, as shown in Fig.~\ref{fig:RFSquidAmp} and described in Eq.~\eqref{eq:rfsquid} for the case of an amplifier composed of RF-SQUIDs. However, this framework for modeling the behavior of these circuits does not consider the intermediate nodes between each pair of RF-SQUIDs, and so ignores the effects of stray capacitances at these nodes which will be present in physical circuits. The introduction of these stray capacitances allows for dynamic instability of the phase at these intermediate nodes, which can excite high-frequency plasma-like modes.

\begin{figure*}
    \centering
    \includegraphics[width = \textwidth]{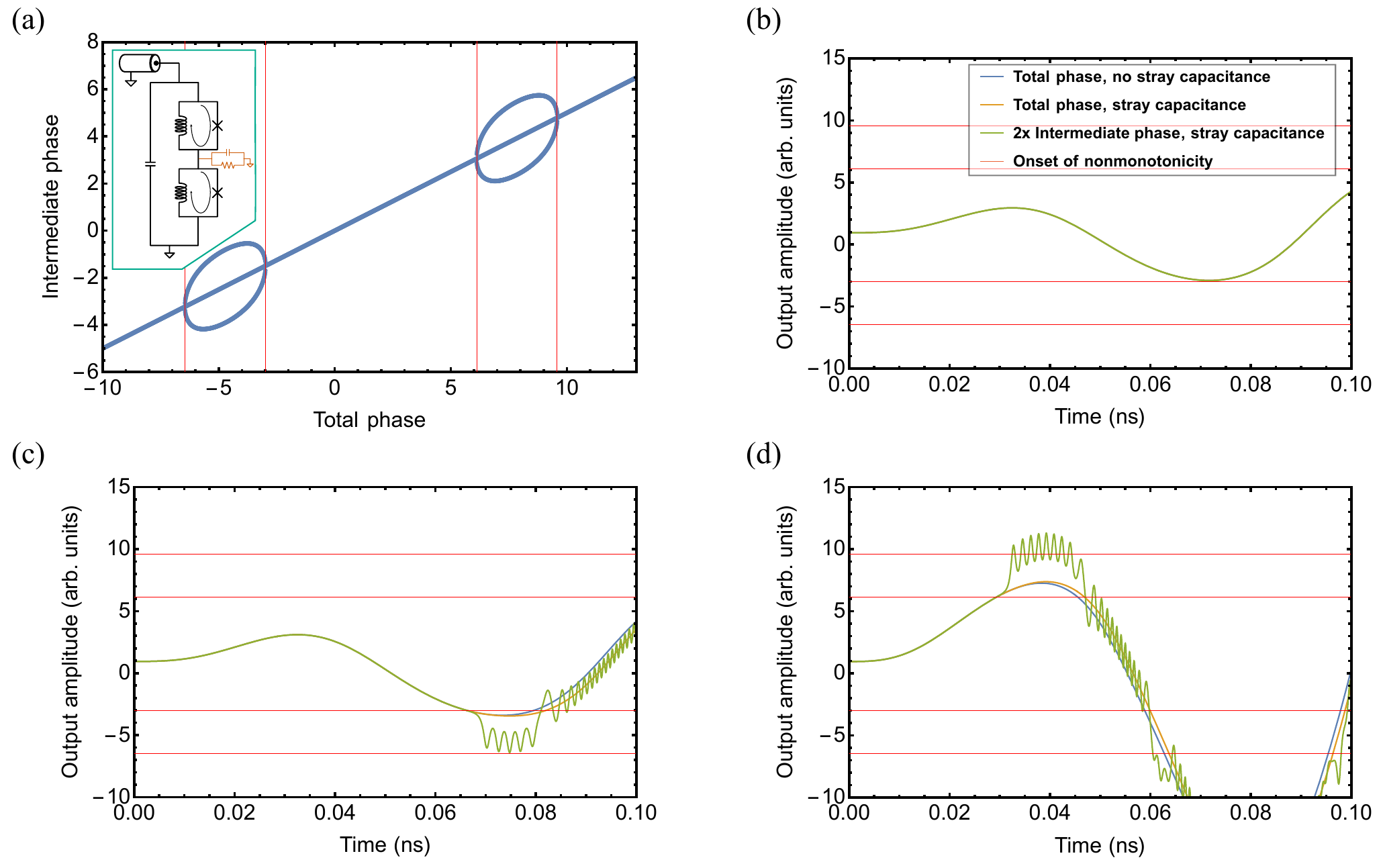}
    \caption{\label{fig:instability} Coupling of low- and high-frequency modes induced by nonmonotonic current-phase relation of constituent RF-SQUIDs. (a) Intermediate phase (i.e. the phase difference across the bottom RF-SQUID, see inset) as a function of the phase difference across both RF-SQUIDs. At the points where the current-phase relation of the RF-SQUID becomes nonmonotonic (indicated by the thin red lines), the solution of the intermediate phase equation trifurcates. Inset: circuit diagram of the amplifier with two RF-SQUIDs. Stray capacitance and damping are indicated in orange. (b-d) Real time dynamics of the amplifier with two RF-SQUIDs depicted in the inset of panel (a) for different pump strength. Panel (b) depicts the weak pump case, where the total phase narrowly avoids the nonmonotonic regime and hence the high-frequency mode of the intermediate phase is not excited. Panel (c) depicts a slightly stronger pump, where the total phase narrowly hits the nonmonotonic regime, which results in the ringing up of the high-frequency mode of the intermediate phase. As the high-frequency mode rings up the dynamic solution with the stray capacitor starts to diverges from the dynamic solution without the stray capacitor. Panel (d) depicts an even stronger pump, where the total phase hits the nonmonotonic regime earlier.}
\end{figure*}

In particular, this problem arises when the current-phase relation of the repeating element in the amplifier is nonmonotonic. In this case, there may be values of the total phase across the amplifier for which multiple divisions of the phase across the $n$-many elements are allowed. With the introduction of dynamic effects at the intermediate nodes between these blocks, we are able to determine how this effect disrupts amplifier performance. To model this behavior, we extend the framework for describing amplifier circuit dynamics in Subsec.~\ref{sec:JPAdesign} by considering intermediate phases $\varphi_i$ between each of the $n$ repeating elements, with $\varphi_1$ representing the phase at the port and $\varphi_{n+1}$ representing the phase below the bottom element, i.e. connected to ground. At the intermediate nodes, we apply a capacitor and resistor in parallel to ground which provide dynamics and damping. This is shown in the inset in panel~(a) of Fig.~\ref{fig:instability}, where we display the two RF-SQUID circuit with an intermediate node. For the case of a circuit with $n$ RF-SQUIDs, we can write the EOM for the intermediate nodes as
\begin{align}
\begin{aligned}
    0=& \left(\frac{\varphi_i-\varphi_{i-1}}{LC}+\frac{i_\text{c}}{\phi_0 C} \sin\left(\varphi_i-\varphi_{i-1}+\varphi_{\text{ext}}\right)\right) \\ &-\left(\frac{\varphi_{i+1}-\varphi_{i}}{LC}+\frac{i_\text{c}}{\phi_0 C} \sin\left(\varphi_{i+1}-\varphi_{i}+\varphi_{\text{ext}}\right)\right) \\ &+\frac{C_s}{C}\ddot{\varphi} +K_s\dot{\varphi_i},
    \label{eq:intermediate}
\end{aligned}
\end{align}
and for the two boundary nodes as
\begin{align}
\begin{split}
    \ddot{\varphi_1}+K\dot{\varphi_1}&+\frac{\varphi_1-\varphi_2}{LC}+\frac{i_\text{c}}{\phi_0 C} \sin\left(\varphi_1-\varphi_2+\varphi_{\text{ext}}\right)\\
    &=2K\dot{\varphi_{\tin}} 
\end{split}
\\
    \varphi_{n+1}&=0
\end{align}
where $C_s$ is the stray capacitance to ground and $K_s$ is the damping rate at the intermediate nodes.

In Fig.~\ref{fig:instability}, we demonstrate how the time-domain output from an amplifier composed of two RF-SQUIDs is disrupted when the current-phase relation is nonmonotonic. As shown in panel~(a), multiple solutions for the phase across each RF-SQUID are allowed when a certain value of the total phase is reached, at which point the solutions for the intermediate phase trifurcate. We see that until the nonmonotonic regime of the RF-SQUIDs is reached, the amplifier behaves as predicted by ignoring intermediate nodes. In order to consider the intermediate nodes, we add a stray capacitance and damping with $C_s=0.001 C$ and $K_s=0.01K$. In panel~(b), we see the time-domain trace resulting from amplifying with a weak pump such that the total phase never reaches the trifurcation point. Here, the total phase as predicted without stray capacitances matches the phase when considering stray capacitances, and the intermediate phase between the RF-SQUIDs is half the total phase as expected. In panel~(c), a stronger pump causes the total phase to pass into the trifurcated regime, and at this point the high-frequency mode of the amplifier is excited. Here, the trace predicted by neglecting stray capacitances drifts from the trace considering stray capacitances. Finally, in panel~(d), the pump is increased further and the trifurcation is reached sooner, again causing the high-frequency mode to be excited as soon as the amplifier enters this regime. Beyond this point, the traces obtained by considering and by ignoring stray capacitances do not agree.

\begin{figure}
    \centering
    \includegraphics[width = 0.9\linewidth]{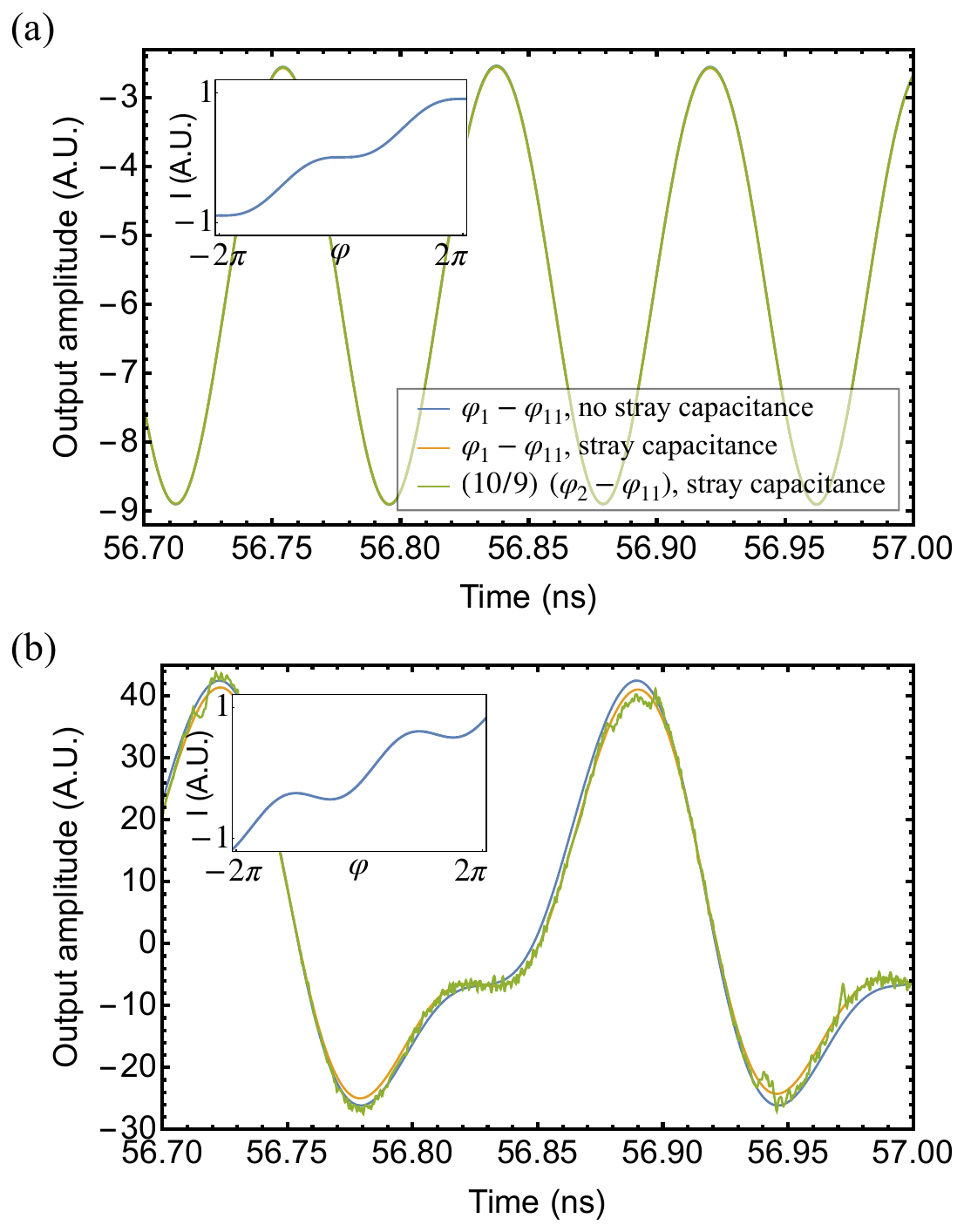}
    \caption{\label{fig:longTimeTraces} Long-time behavior of amplifiers with inductive blocks with differing current-phase relations. A.U. indicates arbitrary units. (a) Dynamics for an amplifier composed of 10 RF-SQUIDs with monotonic current-phase relation (see inset). The time-domain trace indicates that the phase below the first RF-SQUID and above the following 9 is nearly exactly 9/10 the total phase. Or rather, the phase is divided equally among the RF-SQUIDs and the amplifier is well modeled without stray capacitances. (b) Dynamics for an amplifier composed of 10 RF-SQUIDs with nonmonotonic current-phase relation (see inset). The time-domain trace indicates that the phase below the first RF-SQUID and above the following 9 is unstable, since the nonmonotonicity allows for several ways to divide the phase among the RF-SQUIDs. The amplifier is not well modeled when neglecting stray capacitances since the traces do not agree.}
\end{figure}

This behavior persists in the long-time regime for amplifiers with nonmonotonic current-phase relations and many RF-SQUIDs, as shown in Fig.~\ref{fig:longTimeTraces}. We simulate the behavior with an inductive block composed of 10 RF-SQUIDs with dynamics at each intermediate node as specified by Eq.~\eqref{eq:intermediate}. We do this for RF-SQUIDs with both monotonic and nonmonotonic current-phase relations, and we find that in the monotonic case the solution converges to the solution with no stray capacitances. Conversely, in the nonmonotonic case, high-frequency modes are always excited and the solution never converges to the no stray capacitance case. We provide additional traces in short- and long-time regimes for varying degrees of monotonicity in Appendix~\ref{appx:JPA}.

Evidence presented in this section indicates that we should focus on amplifiers with monotonic current-phase relations. We use this as a design rule for amplifier circuits in the remainder of the paper. To enforce this design rule, we follow the prescription of Appendix~\ref{appx:enforce}.

\subsection{Performance of JPA circuits}\label{sec:JPACircuits}

In this subsection we will examine JPA circuits which are optimized to a target gain of $G_\text{t}=20~\dB$, which is a common target in quantum computing applications. We will do so in both the degenerate and nondegenerate cases. In both cases, we optimize with a pump frequency $\omega_{\text{p}}= 12~(2\pi)$~GHz, with $\omega_{\text{s}}=0.5\, \omega_{\text{s}}=6~(2\pi)$~GHz in the degenerate case and $125/249\, \omega_{\text{p}}$ in the nondegenerate case. In optimizing our amplifiers, we treat pump power as a hyperparameter which is fixed for each optimization, and we adjust it to achieve the highest possible PAE. We will first investigate amplifiers whose inductive blocks are composed of RF-SQUIDs, where all circuit parameters as well as the damping rate $K$ are optimized. Then, we investigate amplifier designs with higher PAE by shunting the RF-SQUIDs with three Josephson junctions with an applied current bias. This extended RF-SQUID circuit offers higher PAE at the expense of additional complexity, which may prove impractical to build. In addition to amplifier gain, we provide bandwidth and third order intercept point (IP3) characteristics for our nondegenerate amplifier designs.

\subsubsection{RF-SQUID amplifiers}\label{sec:practical}

We find that by tuning the parameters of an amplifier composed of a chain of identical RF-SQUIDs as in Fig.~\ref{fig:RFSquidAmp}, we can achieve a PAE which is a substantial fraction of the PAE of polynomial amplifiers. In Table~\ref{tab:parameters1}, we list the circuit parameters for our optimized amplifiers. In particular, we tune the Josephson critical current $i_c$, the inductance $L$, the applied flux bias $\Phi_\text{ext}$, and the damping rate $K$. In the degenerate case, we also tune the relative phase offset $\delta$ between the signal and pump.

\begin{figure}
    \centering
    \includegraphics[width = 0.9\linewidth]{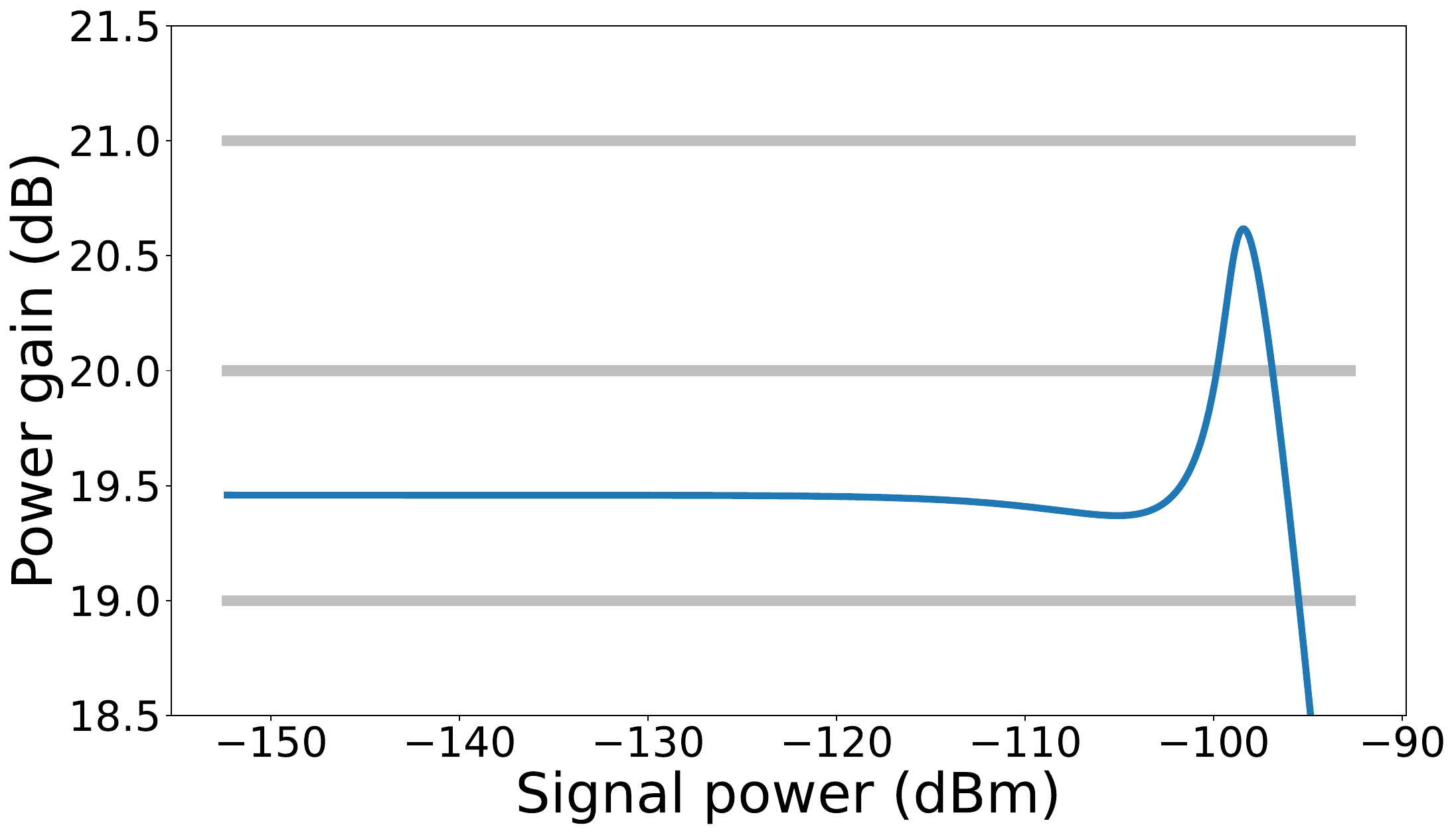}
    \caption{\label{fig:realdp}Gain versus input signal power of the degenerate RF-SQUID amplifier shown in Fig.~\ref{fig:RFSquidAmp}, with optimized circuit tuned to 20~dB gain. Parameters for this amplifier are listed in Tab.~\ref{tab:parameters1}.}
\end{figure}

\begin{figure}
    \centering
    \includegraphics[width = 0.9\linewidth]{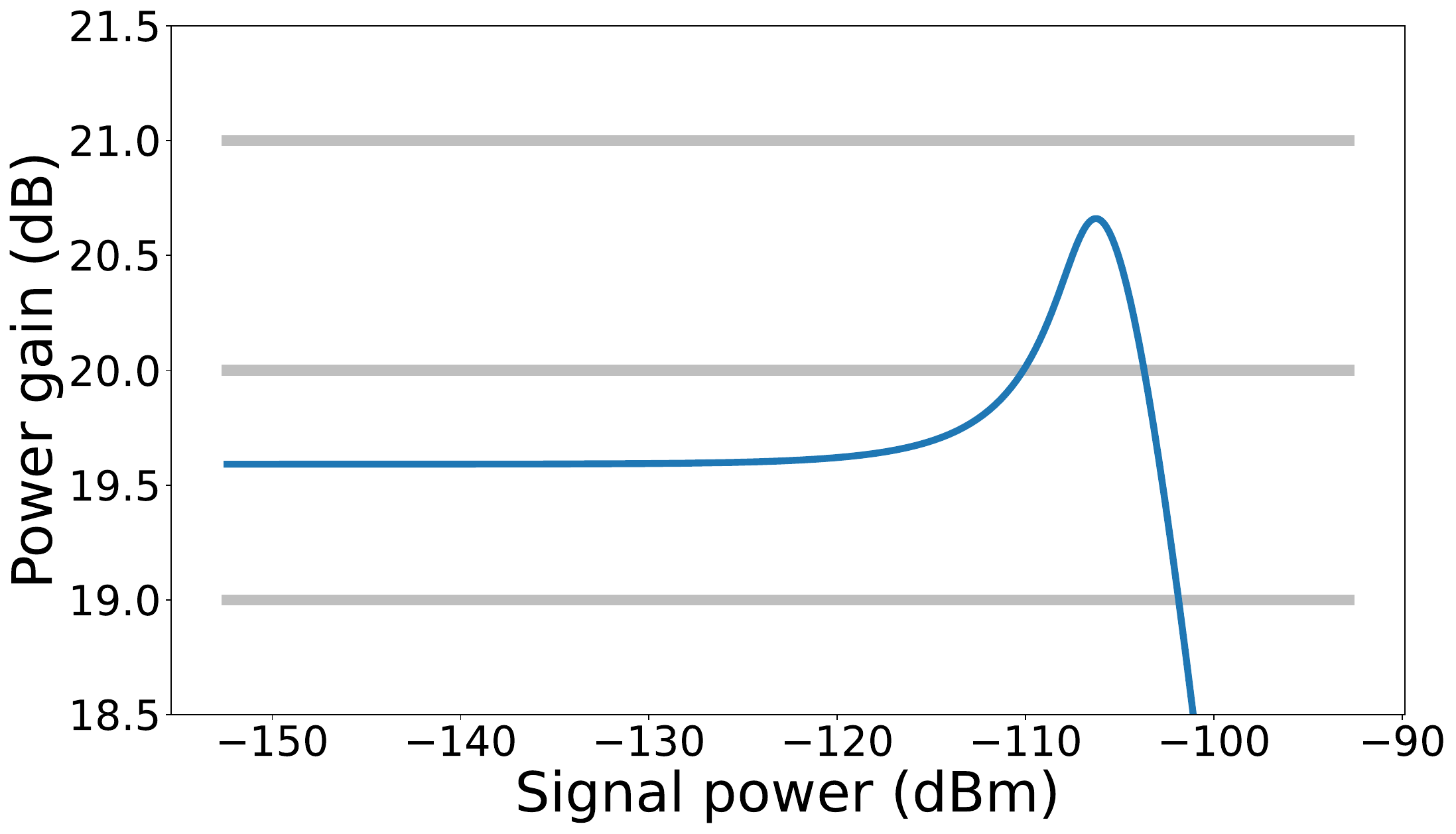}
    \caption{\label{fig:realndp}Gain versus input signal power of the nondegenerate RF-SQUID amplifier shown in Fig.~\ref{fig:RFSquidAmp}, tuned to 20~dB gain. Parameters for this amplifier are listed in Tab.~\ref{tab:parameters1}. The signal frequency is $\omega_\sig=125/249 \, \omega_\pump$. }
\end{figure}

In the degenerate case, we were able to optimize our amplifiers to a PAE of 37.9\%. We plot the gain of this amplifier as a function of signal power in Fig.~\ref{fig:realdp}. In the nondegenerate case, we get a PAE of approximately 8.9\%, with the gain curve shown in Fig.~\ref{fig:realndp}. These circuits have parameters as listed in Tab.~\ref{tab:parameters1}. We note that our optimization is independent of the choice of capacitance $C$, and so our circuit elements, pump power, and characteristic impedance $Z$ can be rescaled by a different choice of $C$. In particular, inductances (e.g., $L$) are inversely proportional to $C$, while currents (e.g., $i_{c}$) and pump power, $P_\text{pump}$, are directly proportional to $C$. Here, we choose $C=0.5$~pF and provide device parameters based on this choice. Our amplifiers are optimized with a low relative pump power (such that $(P_\text{pump})/(n^2 C K \phi_0^2 \omega_\pump^2)$ is small), as we find this improves PAE. Higher output saturation power can be achieved without changing PAE by rescaling parameters as described above.

\subsubsection{Extended RF-SQUID amplifiers}\label{sec:impractical}

\begin{figure}
    \centering
    \includegraphics[width = 0.9\linewidth]{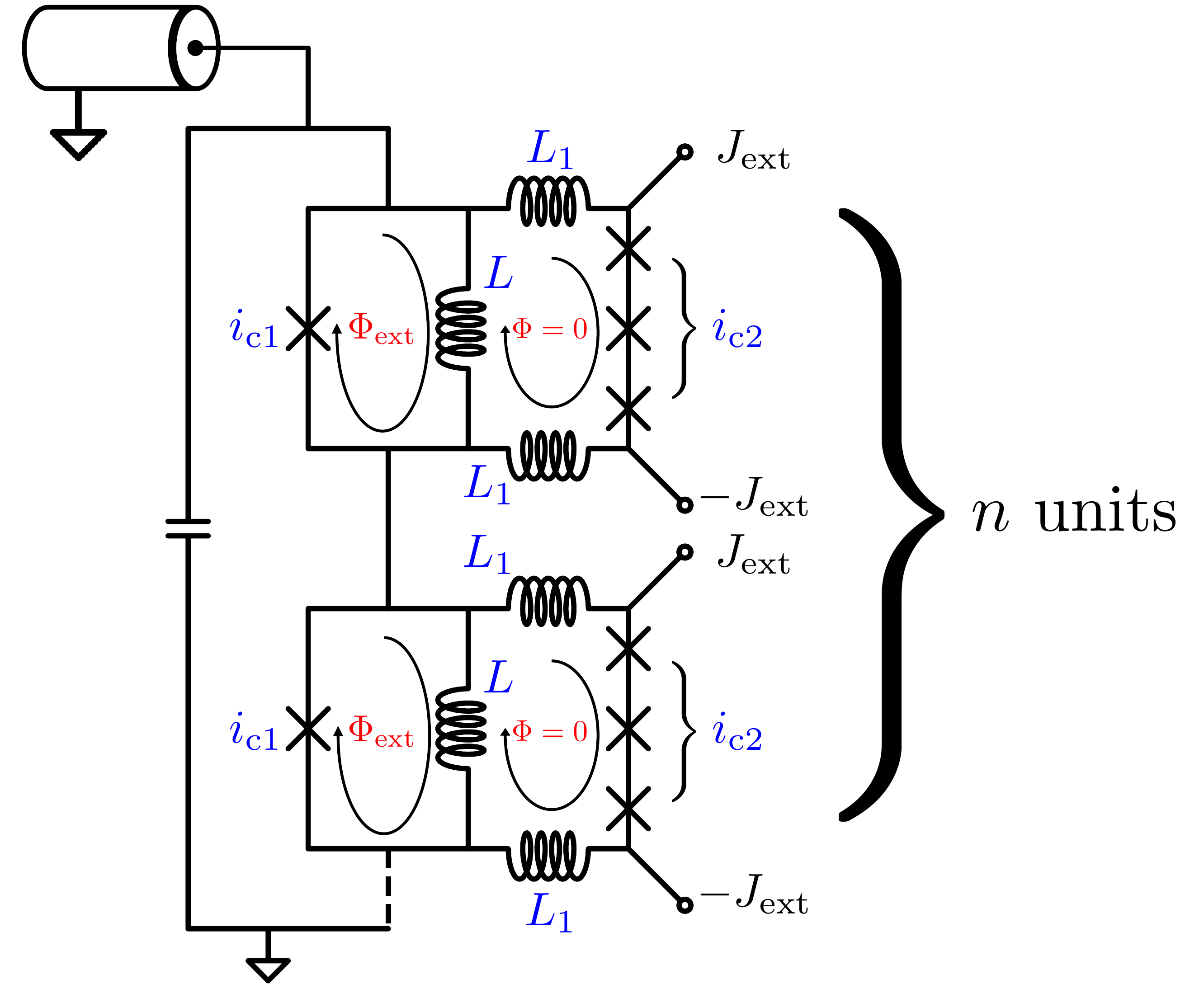}
    \caption{\label{fig:NCurrent}Amplifier circuit design which gives highest PAE. The circuit is composed of $n$-many RF-SQUIDs, each shunted by 3 current-biased Josephson junctions.}
\end{figure}

Here, we investigate amplifiers with inductive blocks composed of RF-SQUIDs shunted by three Josephson junctions in series with an applied current bias, as well as a separate flux bias through the RF-SQUID loop, as shown in Fig.~\ref{fig:NCurrent}. With these extended RF-SQUID designs, we can achieve a PAE which is higher than that of the RF-SQUID amplifiers above by more carefully tuning higher order terms in the current-phase relation of the inductive block of the amplifier.

\begin{figure}
    \centering
    \includegraphics[width = 0.9\linewidth]{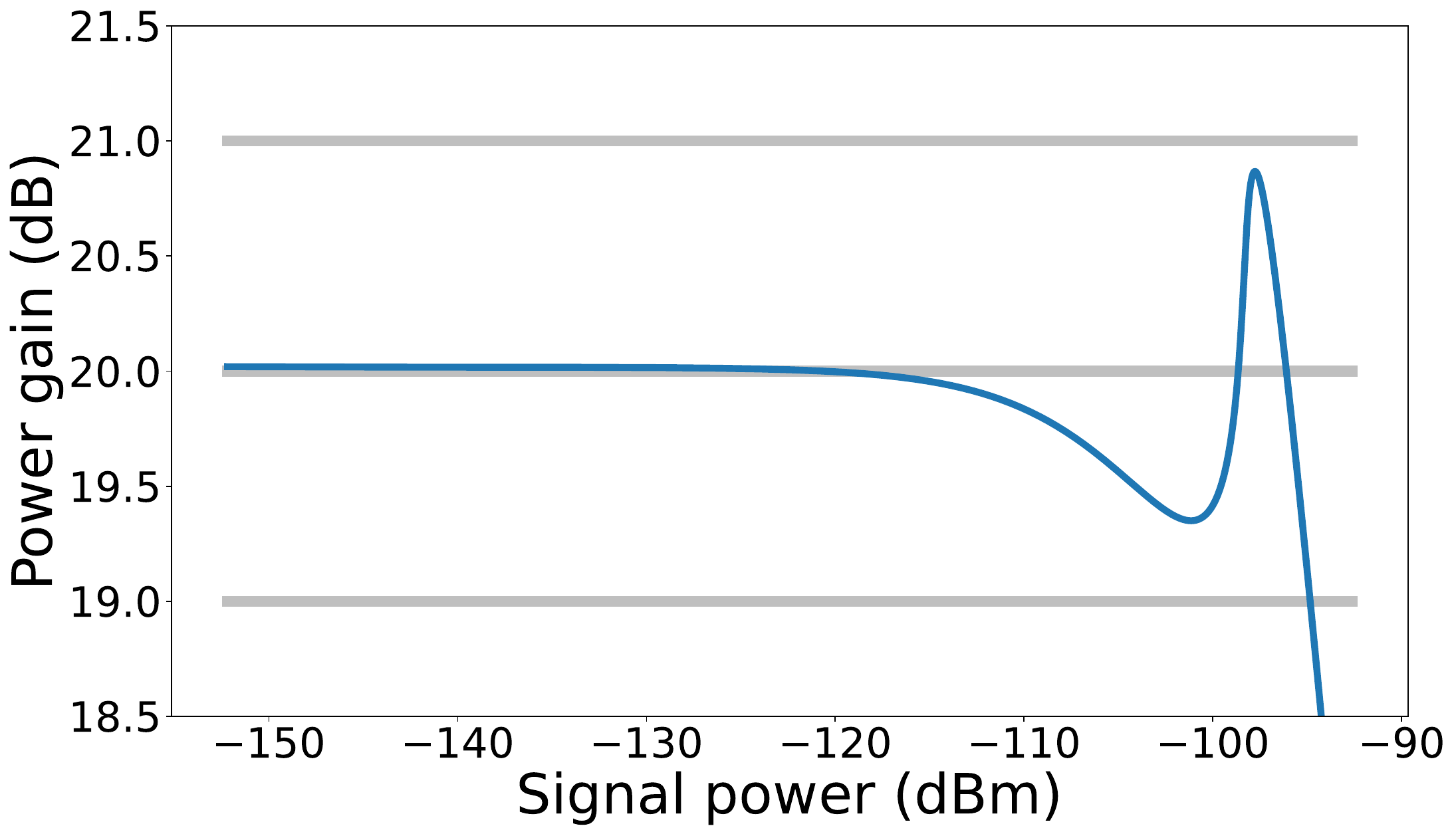}
    \caption{\label{fig:kdp}Gain versus input signal power of the degenerate extended RF-SQUID amplifier shown in Fig.~\ref{fig:NCurrent}, with optimized circuit tuned to 20~dB gain. Parameters for this amplifier are listed in Tab.~\ref{tab:parameters1}.}
\end{figure}

\begin{figure}
    \centering
    \includegraphics[width = 0.9\linewidth]{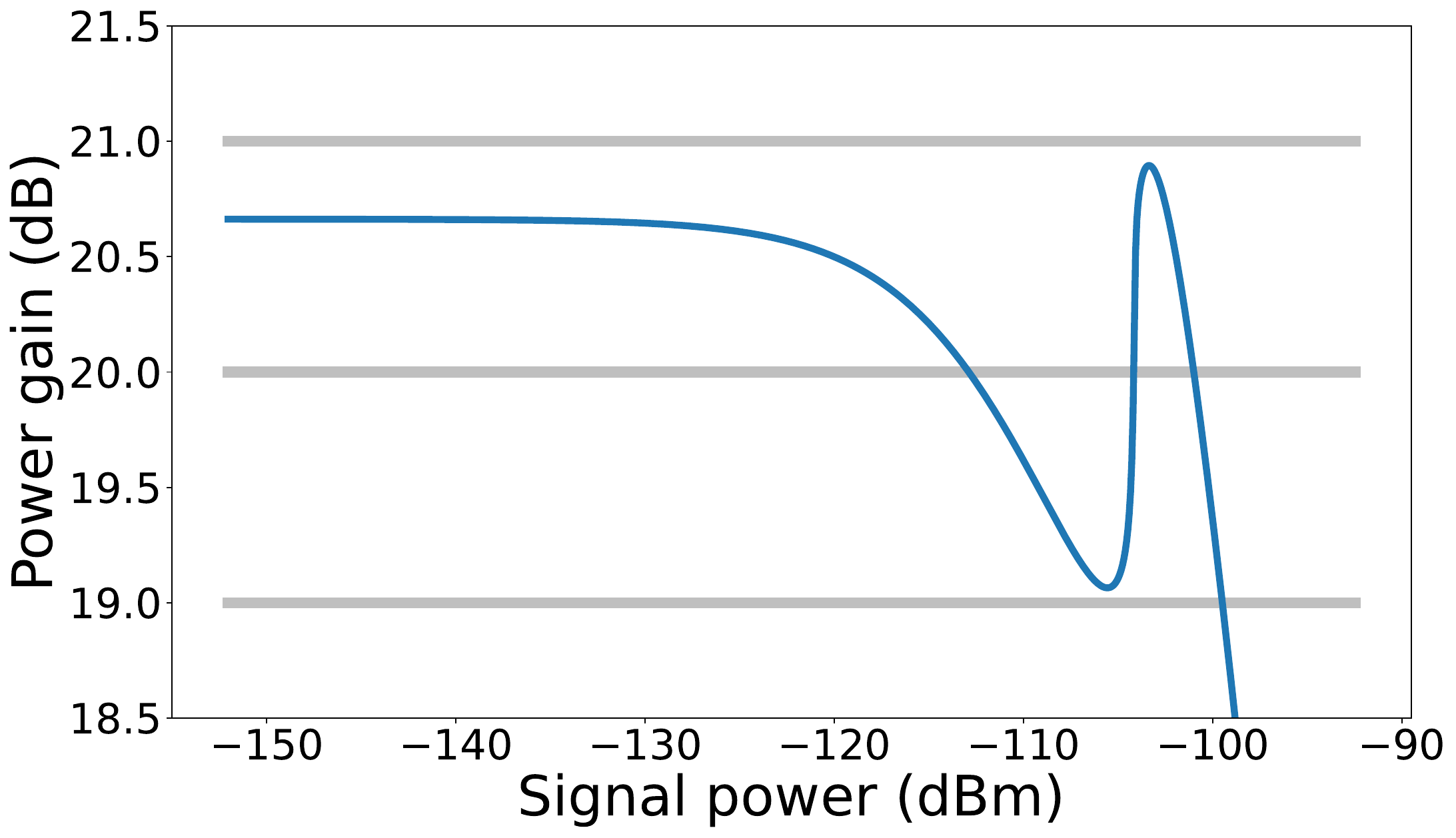}
    \caption{\label{fig:knp}Gain versus input signal power of the nondegenerate extended RF-SQUID amplifier shown in Fig.~\ref{fig:NCurrent}, with optimized circuit tuned to 20~dB gain. Parameters for this amplifier are listed in Tab.~\ref{tab:parameters1}. The signal frequency is $\omega_\sig=125/249 \, \omega_\pump$.}
\end{figure}

Parameter values that optimize this circuit for 20~dB gain can be found in Tab.~\ref{tab:parameters1}. We provide parameters for both the degenerate and nondegenerate cases, with their respective gain curves plotted in Fig.~\ref{fig:kdp} and Fig.~\ref{fig:knp}. The PAE of the degenerate amplifier is approximately 42.6\%, while that of the nondegenerate amplifier is 8.9\%.

\begin{table*}
\begin{ruledtabular}
\begin{tabular}{ccccc}
        parameter & degenerate RF-SQUID & nondegenerate RF-SQUID & degenerate extended RF-SQUID & nondegenerate extended RF-SQUID \\
        \hline
         $P_{\text{pump}}$ & -72.3 dBm & -72.4 dBm & -72.2 dBm & -72.1 dBm  \\
         $n$ & 10 & 10 & 10 & 10 \\
         $K$ & 2.42 $(2\pi)$ GHz & 2.36 $(2\pi)$ GHz & 2.50 $(2\pi)$ GHz & 2.57 $(2\pi)$ GHz \\
         $C$ & 0.5 pF & 0.5 pF & 0.5 pF & 0.5 pF \\
         $Z$ & 131 $\Omega$ & 135 $\Omega$ & 127 $\Omega$ & 124 $\Omega$ \\
         $i_{\text{c}}$ & 18.8 $\mu$A & 18.1 $\mu$A & -- & -- \\
         $i_{\text{c1}}$ & -- & -- & 21.4 $\mu$A & 20.3 $\mu$A \\
         $i_{\text{c2}}$ & -- & -- & -0.921 $\mu$A* & -3.32 $\mu$A* \\
         $L$ & 17.5 pH & 18.1 pH & 15.3 pH & 15.0 pH\\
         $L_1$ & -- & -- & 399 pH & 66.5 pH \\
         $\Phi_{\text{ext}}$ & 3.09 $\phi_0$ & 3.09 $\phi_0$ & 3.03 $\phi_0$ & 3.09 $\phi_0$ \\
         $J_{\text{ext}}$ & -- & -- & 4.65 $\mu$A & 0.133 $\mu$A \\
         $\delta$ & 0.799$\pi$ & -- & 0.876$\pi$ & -- \\
         $PAE$ & 37.9\% & 8.9\% & 42.6\% & 14.2\%
    \end{tabular}
\end{ruledtabular}
\caption{\label{tab:parameters1} Optimized amplifier circuit parameters. Degenerate RF-SQUID and nondegenerate RF-SQUID columns refer to circuits as shown in Fig.~\ref{fig:RFSquidAmp} and described in Subsec.~\ref{sec:practical}, while degenerate extended RF-SQUID and nondegenerate extended RF-SQUID columns refer to circuits as shown in Fig.~\ref{fig:NCurrent} and described in Subsec.~\ref{sec:impractical}.
\\
*Junctions with negative Josephson critical currents should be replaced with DC-SQUIDs with appropriate flux bias which produce this equivalent negative $i_\text{c}$.
}
\end{table*}

\subsubsection{Bandwidth and IP3}

\begin{figure}
    \centering
    \includegraphics[width = 0.9\linewidth]{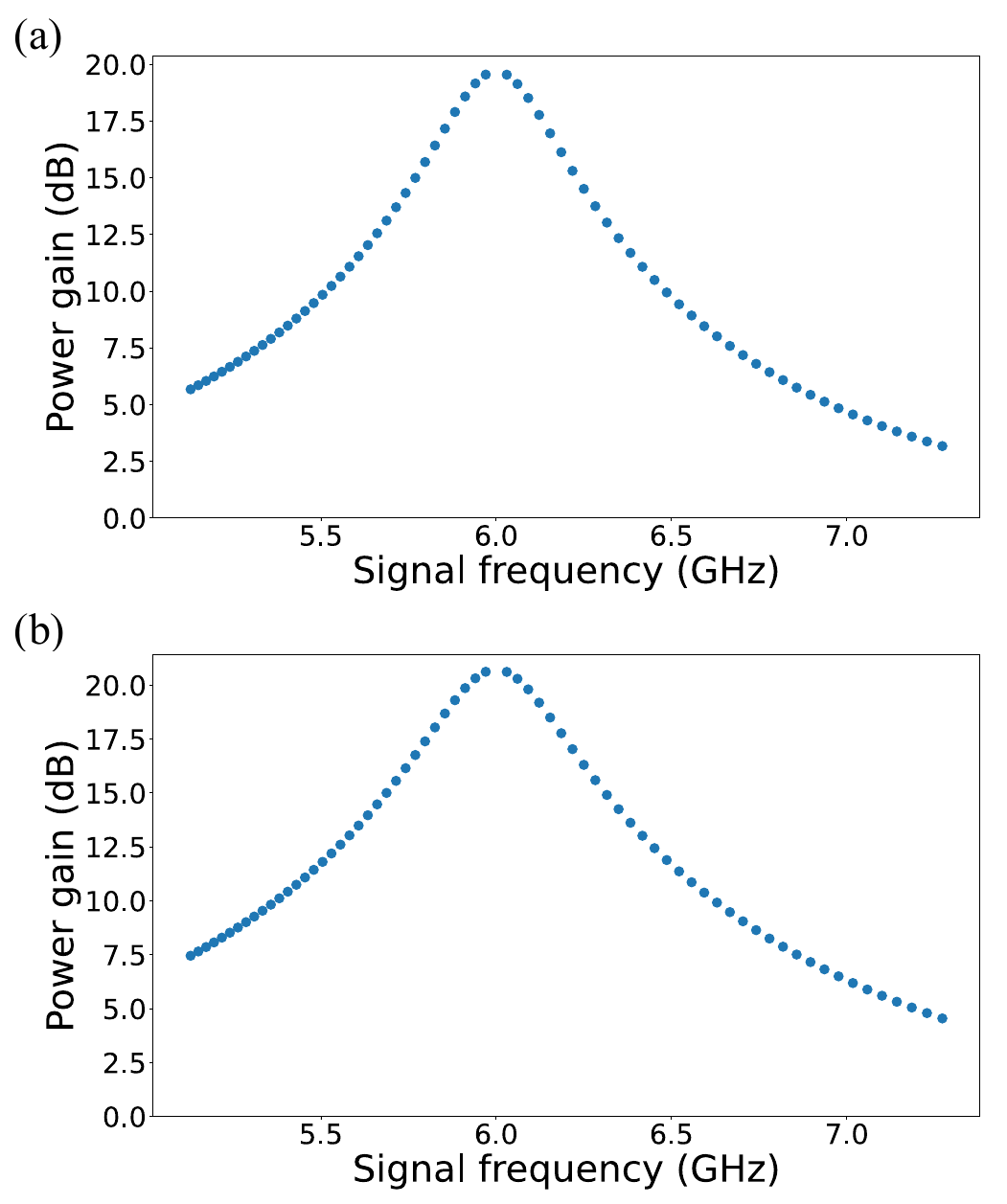}
        \caption{\label{fig:kbw}Amplifier bandwidth: gain of the nondegenerate amplifier versus signal frequency. (a) Amplifier in Subsec.~\ref{sec:practical}. (b) Amplifier in Subsec.~\ref{sec:impractical}.}
\end{figure}

Bandwidth is an important measure of amplifier performance due to the need to amplify signals over a range of frequencies. For the amplifiers in Subsec.~\ref{sec:practical}, we find a bandwidth of 327~MHz. The large bandwidth is a consequence of the high damping rate of about $2.5~(2\pi)$~GHz. We plot the gain of the nondegenerate amplifier over a range of frequencies in Fig.~\ref{fig:kbw}(a), demonstrating this bandwidth. The gain versus signal frequency plot for the nondegenerate amplifier described in Subsec.~\ref{sec:impractical} is shown in Fig.~\ref{fig:kbw}(b), with a bandwidth of 375~MHz.

Another measure of amplifier performance is IP3, which indicates the range of signal power over which intermodulation products between two distinct signal frequencies remain small. Here, we compute the output power of third order intermodulation products resulting from the amplification of two similar but distinct signal frequencies within the amplifier bandwidth, as is specified on pages~511-518 of Ref.~\cite{Pozar}. We note that in order to reduce computation run time, we slightly modify the procedure in Subsec.~\ref{sec:numerics} to use the last half of the DTI solution rather than the last quarter, while ensuring the integration time remains at least 2000 periods of the signal frequencies. This allows us to choose an integration time which is twice $T_\text{min}$ rather than 4 times $T_\text{min}$. This reduced integration time does not affect our solution since for the two similar signal frequencies used in IP3 analysis, the integration time is very long and transient solutions die out far before half the integration time.

\begin{figure}
    \centering
    \includegraphics[width = 0.9\linewidth]{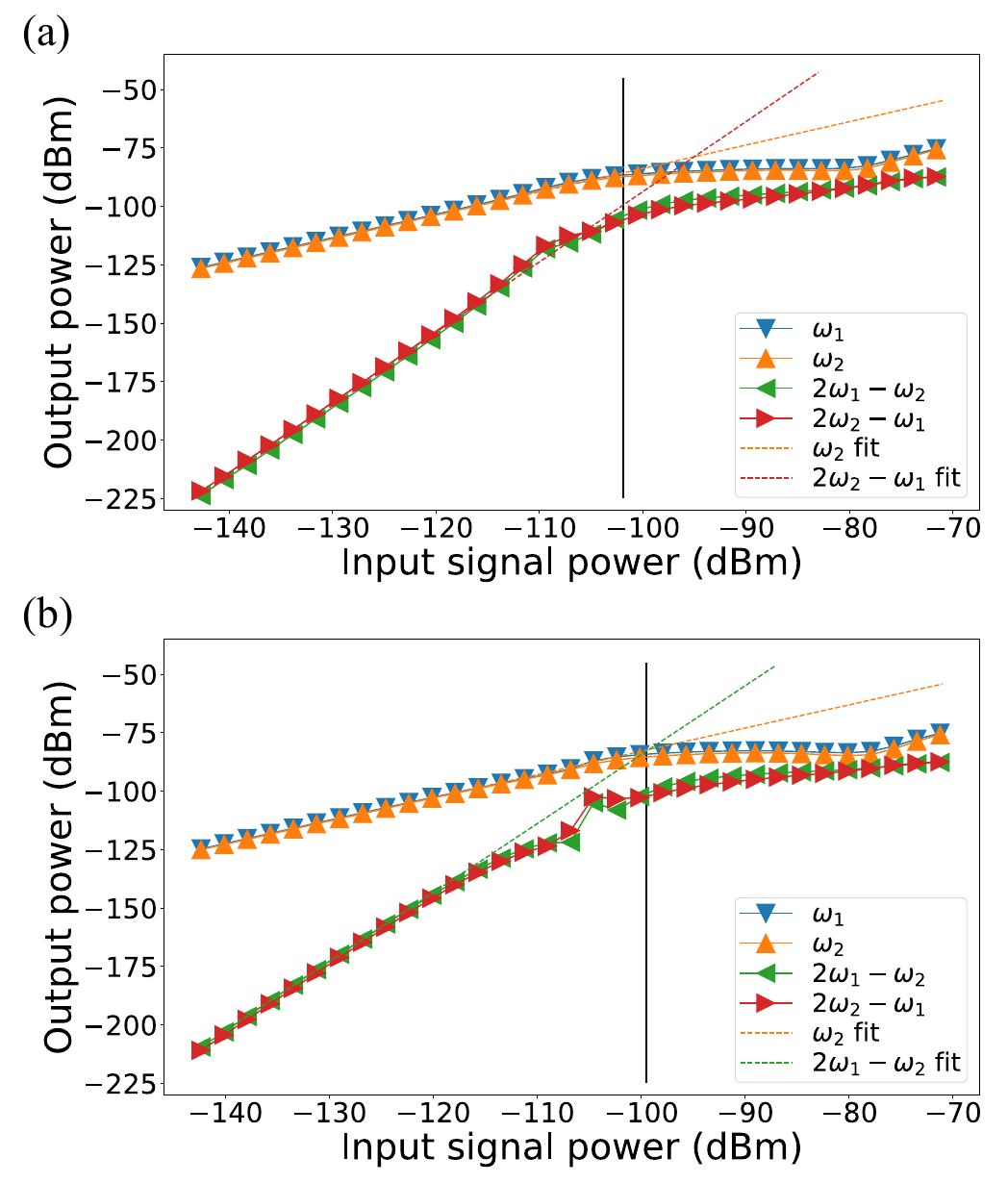}
    \caption{\label{fig:kip3}IP3: output power of signal frequencies as well as their third order intermodulation products, plotted in dBm units. Input signal power refers to the total input power from both signals. Here we have $\omega_1=101/201 \, \omega_\pump$ in blue and $\omega_2=101/200 \, \omega_\pump$ in orange. The third order term $ 2 \omega_1 - \omega_2$ is in green and $ 2 \omega_2 - \omega_1$ is in red. We include fits for $\omega_2$ (dashed orange, which is the lower of the linear components), and $ 2 \omega_2 - \omega_1$ (dashed red, panel~(a), which is the higher of the cubic components) and $ 2 \omega_1 - \omega_2$ (dashed green, panel~(b), which is the higher of the cubic components) which demonstrate the intercept points. These fits take into account the 3rd through 10th points of each data set. The black lines represent the input saturation power as determined by our requirement that the input signal be amplified by 20~dB $\pm$ 1~dB. (a) Amplifier in Subsec.~\ref{sec:practical}. (b) Amplifier in Subsec.~\ref{sec:impractical}.}
\end{figure}

We provide a diagram of the output power of our two signals as well as third order products in Fig.~\ref{fig:kip3}(a), which demonstrates the IP3 performance of the amplifier in Subsec.~\ref{sec:practical}. As anticipated, the output signal power grows linearly with input signal power, while the third order terms grow cubically until the saturation point.
We find input third order intercept point (IIP3) of $-95.0$~dBm and output third order intercept point (OIP3) of $-78.9$~dBm. Similarly, in Fig.~\ref{fig:kip3}(b), we provide the IP3 performance of the amplifier in Subsec.~\ref{sec:impractical}, where we find IIP3 of $-99.2$~dBm and OIP3 of $-82.2$~dBm. In each diagram, the black vertical line represents the input saturation power, demonstrating that the third order intermodulation products remain substantially smaller than the output signals up to this point. For further analysis of intermodulation products and a full power spectrum near the saturation point, see Appendix~\ref{appx:spectrum}.

\subsection{Polynomial amplifier chains}\label{sec:chainAmp}

Polynomial amplifiers always have a higher PAE at relatively low target gains (as shown in Fig.~\ref{fig:etavsGt}). 
Intuitively, without introducing extra higher order nonlinearity, we can obtain a higher PAE by connecting amplifiers with smaller gains as opposed to a single amplifier with higher gain. 
In this section, we first calculate the total PAE for an amplifier chain consisting of two amplifiers. Next, we use the result from Subsec.~\ref{sec:PolyAmpResult} to show that we can increase the total PAE significantly by tuning the gain of the two amplifiers. Finally, we extend our investigation to a chain of $n$ amplifiers that all have the same gain and PAE. These amplifiers are not identical, as $A_{\text{sat}}$ needs to be appropriately scaled to match the signal amplitude in the amplifier chain. By connecting 15 amplifiers in a chain, we can achieve a PAE of 98\% for a 20~dB degenerate amplifier. 

\begin{figure}
    \centering
    \includegraphics[width = \linewidth]{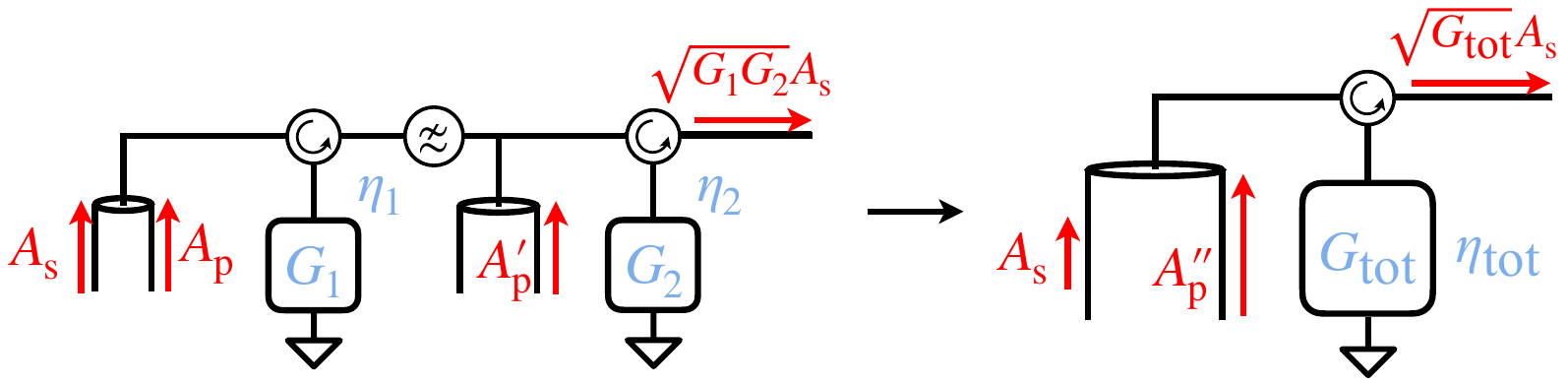}
    \caption{A schematic of an amplifier chain. 
    On the left hand side, we connect two amplifiers with gains of $G_1$ and $G_2$ and PAEs of $\eta_1$ and $\eta_2$, respectively.
    This amplifier chain is equivalent to a single amplifier with total gain $G_\text{tot}$ and achieves a total PAE of $\eta_{\text{tot}}$ shown on the right hand side.
    Perfect circulators and low-pass filters are added to route input, output, and pump waves.
    }
    \label{fig:2amplifiers}
\end{figure}

We start by considering the case where two amplifiers are connected as shown in Fig.~\ref{fig:2amplifiers}. These amplifiers operate at gains of $G_1$ and $G_2$, and achieve PAEs of $\eta_1$ and $\eta_2$, respectively. Let $A_\text{s}$ denote the amplitude of the input signal wave, and let 
$A_\text{p}$ and $A_\text{p}'$ denote the amplitudes of the two pump waves, respectively. The total gain and the total PAE for this amplifier chain, denoted as $G_{\text{tot}}$ and $\eta_{\text{tot}}$, are given by
\begin{align}
    &G_{\text{tot}} = G_1G_2, & \eta_{\text{tot}} =\frac{(G_1G_2-1)\omega_\sig^2A_\sig^2}{\omega_\pump^2{A_\pump'}^2+\omega_\pump^2A_\pump^2}.&
\end{align}
Recalling the definitions of $\eta_1$ and $\eta_2$, 
\begin{align}
      &\eta_1  = \frac{(G_1-1)\omega_\sig^2A_\sig^2}{\omega_\pump^2A_\pump^2}, & \eta_2  =\frac{G_1(G_2-1)\omega_\sig^2A_\sig^2}{\omega_\pump^2{A_\pump'}^2},&
\end{align}
we obtain
\begin{align}
    \begin{aligned}
        \eta_{\text{tot}} =\frac{G_1G_2-1}{\frac{G_1-1}{\eta_1}+\frac{G_1(G_2-1)}{\eta_2}}. \label{eq:etatot}
    \end{aligned}
\end{align}
The PAE of the combined amplifier chain is bounded by the PAE of each individual amplifier,
\begin{align}
\min(\eta_1,\eta_2) \leq \eta_{\text{tot}} \leq \max(\eta_1,\eta_2),
\end{align}
and is determined by how the gain is split between two amplifiers. 
These results can be easily generalized to the $n$-amplifier case, where we have
\begin{align}
   & G_{\text{tot}} = \prod_i G_i,&
     \eta_\text{tot} = \frac{G_{\text{tot}}-1}{\sum_i\frac{G_i-1}{\eta_i}\prod_{k<i}G_i}.&
\end{align}
For a chain of amplifiers that have the same power gain of $G$ and a PAE of $\eta_{\text{PAE}}$, the total gain is $G_{\text{tot}} = G^n$, and the total PAE is simply $\eta_{\text{tot}} = \eta_{\text{PAE}}$. 

\begin{figure}
    \centering
    \includegraphics[width = 0.9\linewidth]{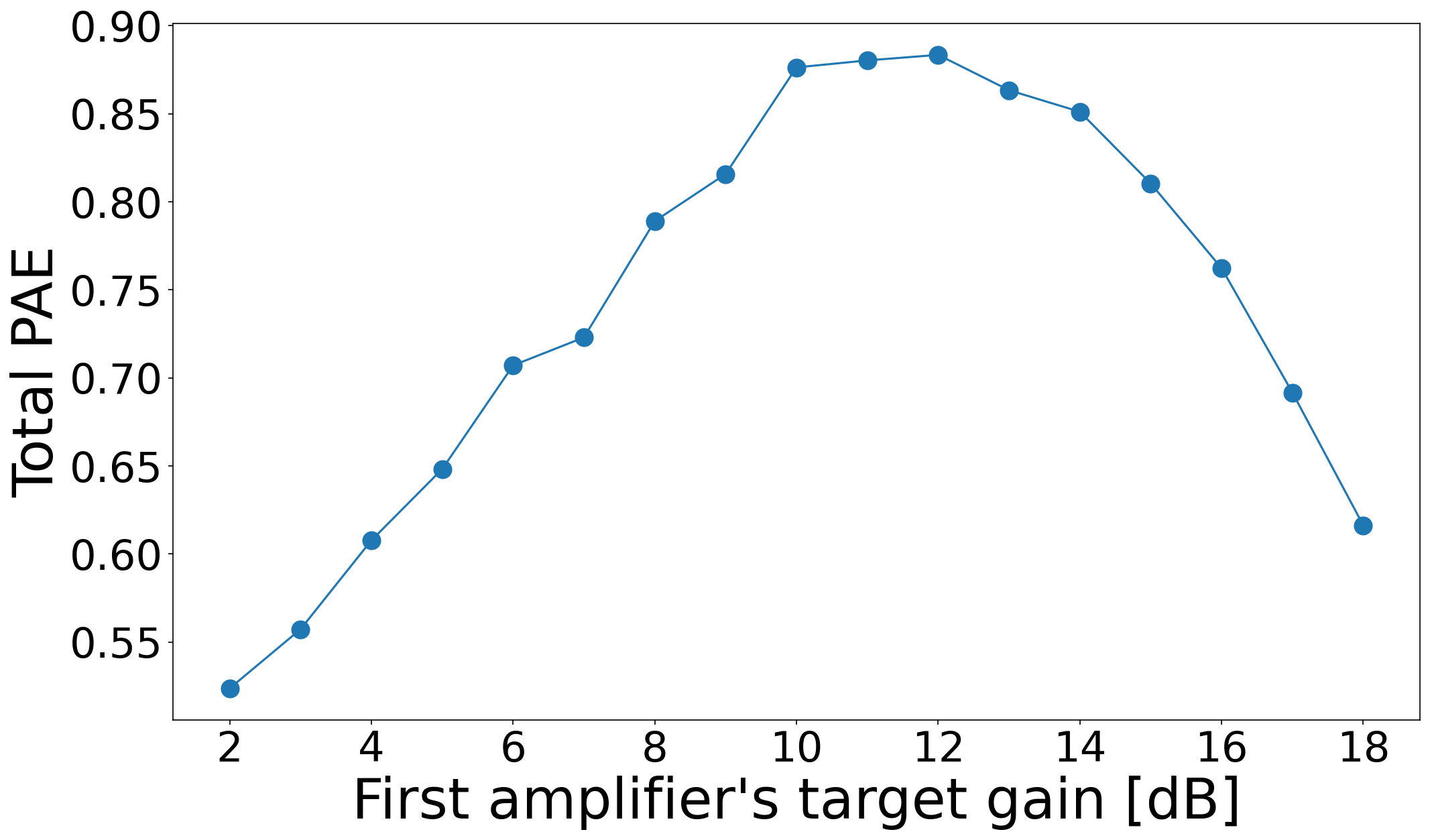}
    \caption{Total PAE of a 20~dB degenerate amplifier chain composed of two 6-th order polynomial amplifiers.}
    \label{fig:twoAmps}
\end{figure}

In Subsec.~\ref{sec:PolyAmpResult}, we optimized a range of $6$-th order polynomial amplifiers designed for various target gains. Specifically, for a target gain of $G_\text{t} = 20$~dB, the highest PAE we obtained was $63.3\%$.
To improve this efficiency, we connect a pair of smaller gain amplifiers to achieve a total gain of 20~dB, where we tolerate variations of up to $\pm 1$~dB from this target. To characterize the amplifier chain, we used the PAE versus gain curve from Fig.~\ref{fig:etavsGt}. To ensure the overall gain variation does not exceed the $\pm 1$~dB tolerance, we made sure that each amplifier in the chain does not vary from its target gain by more than $\pm 0.5$~dB.
The performance of this amplifier chain is illustrated in Fig.~\ref{fig:twoAmps}. We observe that the total PAE of the chain peaks at 88.3\% when the first amplifier is set to a gain of 12~dB and the second to a gain of 8~dB. 

\begin{figure}
    \centering
    \includegraphics[width = 0.9\linewidth]{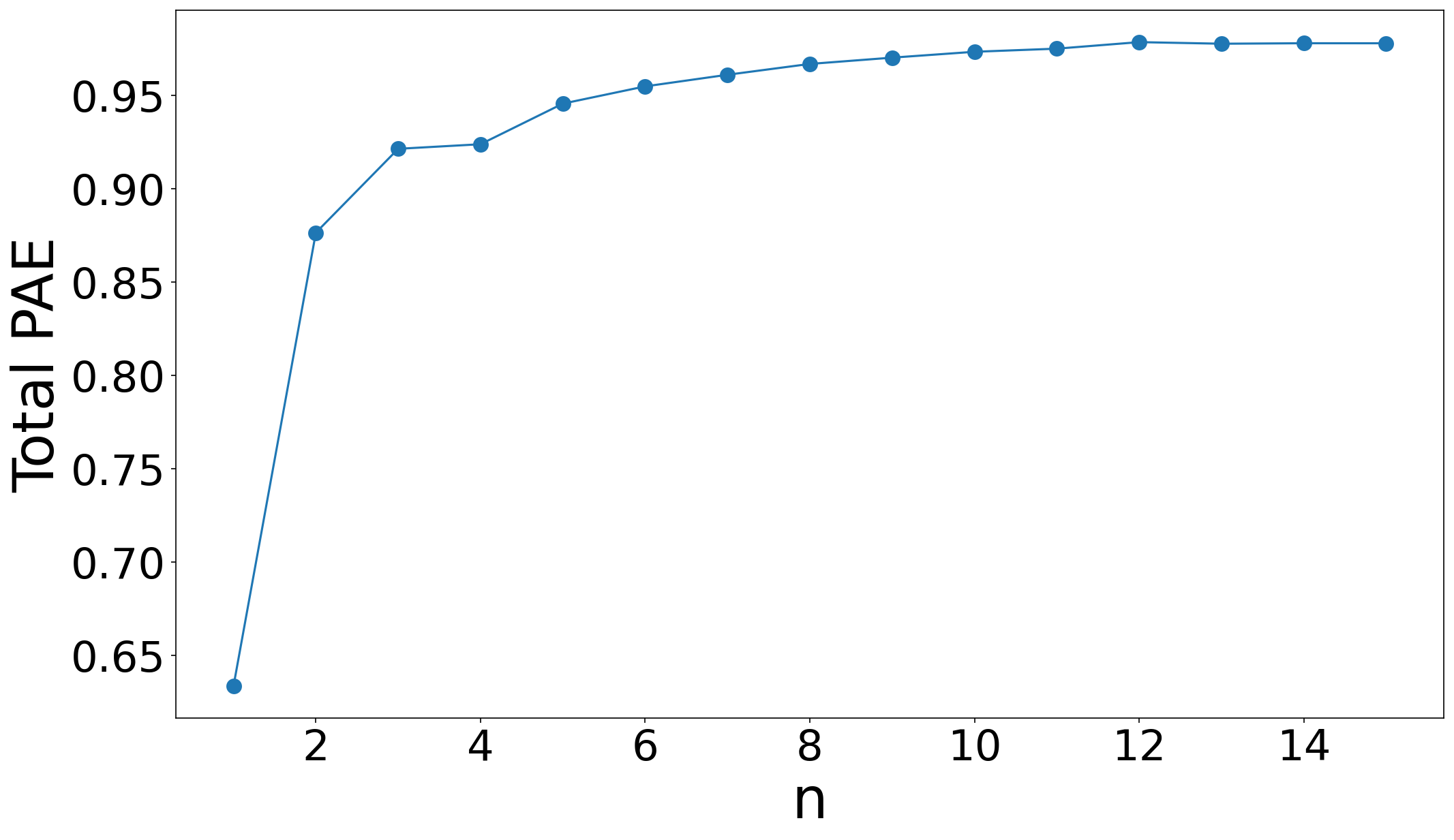}
    \caption{Total PAE of a 20~dB degenerate amplifier chain composed of $n$ same-performance amplifiers.}
    \label{fig:chainAmpsEtavsN}
\end{figure}

We now extend our analysis to a chain of $n$ degenerate amplifiers with each amplifier having the same target gain and PAE. 
We again limit the variation of the total gain to $\pm 1$~dB by tolerating a variation of up to $\pm 1/n$~dB for each amplifier.
As we increase the number of amplifiers from 1 to 15, the total PAE gradually increases from$~63\%$ to$~98\%$, as presented in Fig.~\ref{fig:chainAmpsEtavsN}.  Extrapolating from Fig.~\ref{fig:etavsGt}, we anticipate that the total PAE would eventually approach 100\% as $n\! \to\! \infty$.

\section{Discussion}

Due to the limited power added efficiency of modern JPAs and the need for low noise, efficient amplifiers in quantum computing applications, we have explored optimizing JPA designs by finely tuning the amplifier's inductive block. We find that with the freedom to define an arbitrary inductive block, we can design amplifiers with a PAE orders of magnitude higher than that of modern JPAs. In particular, modern JPAs currently in use are tuned so as to optimize third order nonlinearity. These intrinsic third order terms dynamically generate higher order nonlinearities, which shift signal and idler modes off resonance~\cite{liu2020JRM}. For example, the $g_3$ term in Eq.~\eqref{eq:polyEOM} generates a response at zero frequency and twice the signal frequency which is quadratic in signal amplitude. This in turn generates a cubic response at the signal frequency, which looks like the $g_4$ term of Eq.~\eqref{eq:polyEOM}. Here, we counter this effect by tuning intrinsic higher order nonlinearities in order to cancel the dynamically generated nonlinearities. Using polynomial amplifier designs, we find that we can reach a PAE of around 63\% at 20~dB gain, which is the upper limit in the nondegenerate case.

Restricting ourselves to inductive blocks composed of circuit elements, we identify a rule for designing JPAs. In particular, we find that the inductive block should have a monotonic current-phase relation to avoid issues with how phase divides among elements in the block. Further, considering inductive blocks consisting of circuits extending previous RF-SQUID designs with monotonic current-phase relations, we still realize a significant improvement in PAE. We are able to design JPA circuits with high PAE, large bandwidth, and limited intermodulation distortion. In particular, we provide designs for amplifiers based on chains of RF-SQUIDs with high PAE, and provide designs with chains of extended RF-SQUIDs with even higher PAE at the expense of manufacturing complexity. In Appendix~\ref{appx:variations}, we discuss the robustness of these designs to small manufacturing imperfections.

These amplifiers could prove to be useful in quantum computing applications which require the amplifier to be located in proximity to qubits at millikelvin temperatures, where minimizing heat dissipation is a priority. Compared to previous JPA designs, the amplifiers in this paper are able to provide similar output signal power while requiring a substantially weaker pump, thereby reducing heat load. These high saturation power and high PAE amplifiers may be particularly advantageous in scenarios where simultaneous multiplexed qubit readout is necessary\footnote{Our amplifiers continue to demonstrate good PAE when amplifying multiple distinct signals. Appendix~\ref{appx:spectrum} addresses the power spectrum for two signals near the saturation point.}, as the power requirement scales with the number of qubits. Additionally, these amplifiers demonstrate low distortion at low signal power, which provides a significant benefit in using high saturation power amplifiers as compared to low saturation power amplifiers~\cite{Boutin17,sivak2019}. We anticipate many, or at least several, of these devices to be mounted in a single dilution refrigerator. Recent implementations of low PAE, high saturation power amplifiers are problematic~\cite{Parker2022threewavemixing, Kaufman2024} due to the extremely high pump powers utilized in these experiments, which in turn generate excessive heat in the cryostat.

Finally, by considering chains of polynomial amplifiers, we explore the maximum possible PAE for degenerate amplification. We note that by increasing the number of amplifiers in our chain, we are able to achieve a higher PAE which tends to 100\%. We believe that the advances in theoretically achievable PAE which we outline here can serve as a basis for future parametric amplifiers with use cases where reduced power consumption and dissipation are essential.

\begin{acknowledgments}
We thank Chenxu Liu and Jos\'e Aumentado for useful discussions.
This research was primarily sponsored by the Army Research Office and was accomplished under Award Numbers W911NF-18-1-0144 (HiPS) and W911NF-23-1-0252 (FastCARS). The views and conclusions contained in this document are those of the authors and should not be interpreted as representing the official policies, either expressed or implied, of the Army Research Office or the U.S. Government. The U.S. Government is authorized to reproduce and distribute reprints for Government purposes notwithstanding any copyright notation herein.
ZL and RM acknowledge partial support from the National Science Foundation under award No.~DMR-1848336 and BM acknowledges support from the NSF Graduate Research Fellowship Program.  ZL additionally acknowledges partial support from the Pittsburgh Quantum Institute.
\end{acknowledgments}

\bibliography{jpaTheory}

\clearpage

\onecolumngrid
\appendix

\section{Additional tables}

\begin{table*}[h!]
\begin{ruledtabular}
\begin{tabular}{cl}
        symbol & meaning \\
        \hline
         $\phi$ & node flux of the amplifier \\
         $\phi_0=\hbar/2e$ & reduced magnetic flux quantum \\
         $\varphi=\phi/\phi_0$ & dimensionless node flux of the amplifier \\
          $\phi_{\tin}, \phi_{\out}$ & incoming and outgoing waves on transmission line \\
          $\Phi_\text{ext}$ & external magnetic flux \\
         $C_l,L_l$ &  capacitance and inductance per unit length of the transmission line\\
         $Z = \sqrt{L_l/C_l}$ & characteristic impedance of the transmission line\\
         $\mathcal{L}_\mathrm{s}, \mathcal{L}_\text{tl}$ & Lagrangians for the amplifier and transmission line \\
         $E[\phi]$ & energy of the inductive block of the amplifier
         \\
         $J[\phi] = {\mathrm{d} E[\phi]}/{\mathrm{d}\phi}$  & current-phase relation of the inductive device
         \\
         $A_\text{s},A_\text{p}$ & incoming signal/pump wave amplitudes \\
         $\omega_\text{s},\omega_\text{p}$&incoming signal/pump wave frequencies\\
         $\delta$ &  phase difference between the signal and pump waves\\
        $K$ & damping rate of the amplifier \\
        $G_\text{t}$ & target gain\\
        $A_{\text{sat}}$& signal amplitude at saturation point \\
        
        $\eta_\text{PAE}$ & power added efficiency \\
    \end{tabular}
\end{ruledtabular}
 \caption{\label{tab:symbol} List of commonly used symbols.}
\end{table*}

\begin{table*}[h!]
\begin{ruledtabular}
\begin{tabular}{ccccccccccc}
        order &$\omega_0 $& $K$ & $g_3$ & $g_4$ & $g_5$ & $g_6$&$g_7$&$g_8$&$g_9$ & $g_{10}$ \\
        \hline
         4 &1.4661& $ 0.6879$& $0.7307$&$0.1397$&$-$&$-$&$-$&$-$&$-$&$-$\\
         6&1.5444&$0.8642$&$ 1.1221$& $1.2617$& $0.9072$& $0.2224$&$-$&$-$&$-$&$-$\\
         8&1.6969&$0.9797$& $0.5634$& $-0.04299$& $-0.02057$&$ -0.1392$& $-0.04461$& $0.01779$&$-$&$-$\\
         10&1.6889&$1.0151$& $ 0.6215$& $0.02882$& $0.3609$&$ -0.2192$& $-0.00068101$& $0.0003138$& $0.0003870$&$0.003761$
    \end{tabular}
\end{ruledtabular}
\caption{\label{tab:optimization} Optimization parameters for 20~dB polynomial degenerate amplifiers with different orders of nonlinearity.}
\end{table*}

\section{Harmonic balance method}\label{appx:HBmethod}

In this section, we present the technical details of the HB method for solving the following differential equation:
\begin{align}
    \ddot{\phi} + K\dot{\phi} +{J[\phi]}=2K\dot{\phi}_{\tin},\label{eq:diff}
\end{align}
with the incoming wave $\phi_{\tin}\! = \!A_{\text{s}} \sin({\omega_\text{s}}t) + A_{\text{p}} \sin({\omega_\text{p}} t+\delta)$.

We start with the degenerate polynomial amplifier, where $\omega_\pump = 2\omega_\text{s}$.  In this case, we propose that the solution takes the following form:
\begin{align}
    \phi(t) = a_0 + a_{1}e^{i\omega_\sig t} + \cdots + a_{n}e^{in\omega_\sig t} + h.c..
\end{align}
We denote the Fourier coefficients by the vector $\mathbf{v} = (a_0, a_1, \cdots, a_{n})^{T}$. Using this notation, the differential equation~\eqref{eq:diff} can be rewritten as:
\begin{align}
    (\hat{D}^2+ K\hat{D})\mathbf{v} + \mathcal{J}(\mathbf{v}) = 2K \hat{D} \mathbf{v}_{\tin}, \label{eq:alge}
\end{align}
where $\hat{D}$ is the differential operator, represented by the matrix $\hat{D}_{ab} = ia\omega_\sig\delta_{ab}$; $\mathbf{v}_{\tin}$ denotes the Fourier coefficients of the incoming wave; 
and $\mathcal{J}(\mathbf{v})$ is the nonlinear function in the frequency domain obtained via a Fourier transform of $J(\phi(t))$. The explicit form of $\mathcal{J}(\mathbf{v})$ is hard to find. Thus, we calculate $\mathcal{J}(\mathbf{v})$ using the Fourier transform,
\begin{align}
    \mathbf{v} \xrightarrow{\text{invFT}} \phi \xrightarrow{J} J[\phi] \xrightarrow{\text{FT}} \mathcal{J}(\mathbf{v}).\label{eq:FFT}
\end{align}

To solve this nonlinear algebraic equation~\eqref{eq:alge}, we use the gradient descent algorithm to minimize the following cost function,
\begin{align}
    \mathcal{E}(\mathbf{v}) = \big| (\hat{D}^2+ K\hat{D})\mathbf{v} + \mathcal{J}(\mathbf{v})- 2K \hat{D} \mathbf{v}_{\tin}\big|.
\end{align}
A good initial guess to the solution speeds up the algorithm drastically. One suitable initial guess is the solution to the linear part of the equation, which is given by $\mathbf{v}_0 = 2K(\hat{D}^2+ K\hat{D})^{-1} \hat{D}\mathbf{v}_{\tin}$. Another appropriate initial guess is the solution to the equation with a similar condition. For instance, the solution $\mathbf{v}_1$ to Eq.~\eqref{eq:alge} with the incoming wave $\phi_{\tin} = \phi_{\tin,1}$ can serve as a good initial guess for the same equation with the incoming wave $\phi_{\tin,2}$ which is not substantially different from $\phi_{\tin,1}$.

The situation becomes more complicated in the nondegenerate case, where $\omega_\pump = \omega_\sig + \omega_\text{i}$ and $\delta \omega = (\omega_\sig - \omega_\text{i})/2 \neq 0$. 
In this case, we need to add more harmonics in our proposed solutions. When the signal amplitude is much smaller than the pump amplitude, we only need to add the single signal photon process, i.e., the term with a frequency of $N\omega_\pump \pm \omega_\sig$. Then, we can use the same techniques to solve this nondegenerate amplifier. However, we need to be cautious that, in the Fourier transform~\eqref{eq:FFT}, we apply proper zero-padding since we only keep the single signal photon process. Thus, the HB method is usually slower in the nondegenerate case. When the amplitude of the signal is comparable to that of the pump, the algorithm's performance deteriorates as the higher order photon processes we neglect become significant. In principle, we could include additional components in the solution, but this will considerably slow down the HB algorithm.

\section{Avoiding local minima in energy-phase relations and enforcing monotonicity in current-phase relations}\label{appx:enforce}

In this work, we found that certain amplifier designs produce unstable amplification. As a result, we consider two different limitations on amplifier design which we must enforce when optimizing our amplifiers. First, for all amplifiers of the form in Fig.~\ref{fig:HCircuit}, it is crucial that $\phi = 0$ is the unique minimum of the energy function $E[\phi]$. This condition ensures that the amplifier does not jump between different local minima as the signal amplitude increases, which could result in unreliable amplification. Since current is the derivative of energy with respect to phase, we impose this condition by requiring that the current is strictly positive for $\phi > 0$ and strictly negative for $\phi < 0$.
Second, amplifiers with inductive blocks with identical repeating elements must have monotonic current-phase relations, as discussed in Subsec.~\ref{sec:modeCouple}. This requirement is stricter than the first requirement.

To enforce these requirements while optimizing our amplifiers, we introduce a penalty term in the cost function to discourage designs that violate these constraints. Since it is impractical to track all current values during the optimization process, we instead generate a discrete set of points from the current-phase relation within the range $[\Phi_-, \Phi_+]$ with an appropriate density. This phase range must be sufficiently large to account for fluctuations in $\phi(t)$ throughout the process. For polynomial amplifiers, to enforce that the current  is strictly negative for $\phi < 0$ and strictly positive for $\phi > 0$, we define the following penalty function:
\begin{align}
    \text{penalty} = \lambda \big( \max(0,I^{\max}_-) + \max(0,-I^{\min}_+)\big), 
\end{align}
where $\lambda$ is the regularization factor, and 
\begin{align}
    &I^{\max}_- = \max\big( \{I(\phi)| \Phi_- \leq \phi < 0 \}\big), &I^{\min}_+ = \min\big(\{I(\phi)| 0 < \phi 
    \leq\Phi_+ \}\big).&
\end{align}
For inductive blocks composed of identical repeating elements such as our extended RF-SQUIDs, we check that the current increases between any two successive points with increasing phase, guaranteeing monotonicity. In this case, for each extended RF-SQUID, we consider a grid of 100 points for phase values between $-6\pi$ and $6\pi$. The penalty which enforces this is
\begin{align}
    \text{penalty} = \lambda \big(
    \tanh (-\Delta\text{I}_\text{min}-0.1)+1
    \big), 
\end{align}
where $\Delta\text{I}_\text{min}$ is the minimum difference between sequential current values in the grid.

In the case of an inductive block with repeating RF-SQUIDs, we impose this requirement analytically, which is more computationally efficient. Considering the EOM in Eq.~\eqref{eq:rfsquid}, the current-phase relation is monotonic when the oscillations of the $\sin$ term are smaller than the slope of the linear term. That is, we enforce that $i_\text{c}L/\phi_0<1$. In this case, the penalty is
\begin{align}
    \text{penalty} = \lambda \big(
    \tanh (~20~(i_\text{c}L/\phi_0-1.15))+1
    \big). 
\end{align}

For JPA circuits, we also find that it can be useful to use an alternative cost function for finer optimization from a high-PAE starting point. In these cases, we use the cost function
\begin{align}
\begin{aligned}
f\left(\omega_0,K,\cdots\right) =
          \sum_{m=1}^{N} \exp{ \left( -1500/ \left(G\left(A_\sig = \frac{mR}{N} ,\omega_0,K,\cdots\right) - G_\text{t} \right)^2 \right)} + \text{penalty}
\end{aligned}
\end{align}
instead of the cost function specified in Eq.~\eqref{eq:CostFunction}. Here, if $G=G_t$ for any value of $m$, then the cost for that point is 0.

\section{Numerical solution and optimization of JPA circuits}\label{appx:JPA}

\begin{figure}
    \centering
    \includegraphics[width = \textwidth]{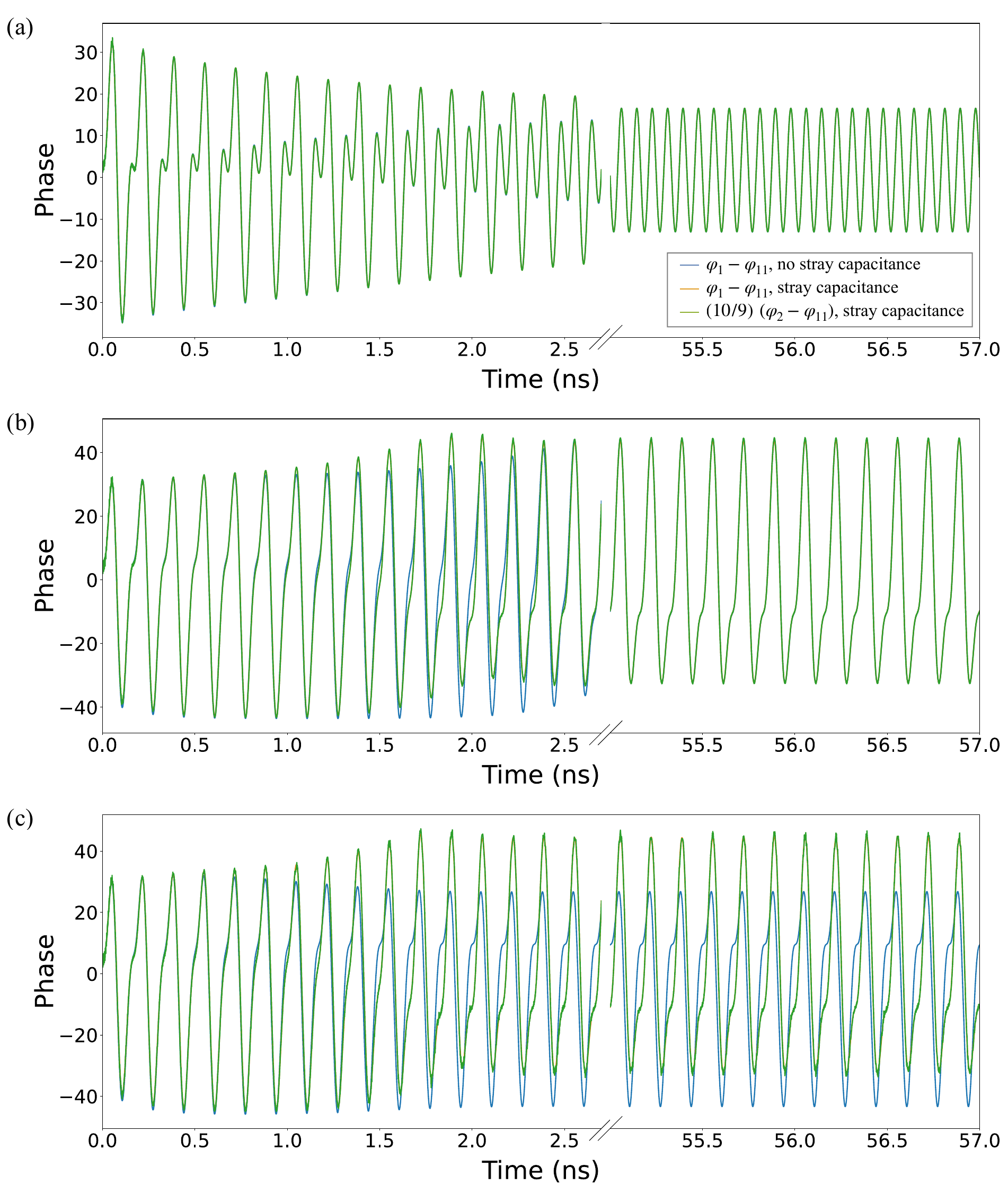}
    \caption{Example time-domain output signals computed by DTI for amplifiers with 10 RF-SQUIDs. In each panel, we plot the signal for short-time, demonstrating the transient behavior, and for long-time, demonstrating the steady-state behavior. (a) Amplifier with inductive block whose current-phase relation is very monotonic, i.e. its slope is never near zero. In this case, we see the initial transient behavior, and then the steady state behavior, with agreement among simulations with and without stray capacitances, and at intermediate phases. (b) Amplifier with inductive block whose current-phase relation is nearly nonmonotonic, i.e. it is monotonic but its slope nears zero. Here, we see transient behavior, but the steady state settles with agreement among simulations. (c) Amplifier with inductive block whose current-phase relation is nonmonotonic. In this case, the amplifier never settles, behavior is not periodic even in the long-time regime, and the simulations do not agree.}
    \label{fig:transient}
\end{figure}

In solving the EOMs for a circuit of the form described in Fig.~\ref{fig:HCircuit}, we use the DTI method described in Section~\ref{sec:numerics} to compute the output signal from the amplifier. We use a total integration time that is commensurate with the period of the pump and signal frequencies. That is, we enforce that the integration time must be a multiple of $T_{\min}=\text{lcm}(\frac{2\pi}{\omega_\pump},\frac{2\pi}{\omega_\sig})$. We choose an integration time of $4 n_c T_\text{min}$ since we wish to isolate the steady-state solution to the equations of motion, while at time near $t=0$ we see a transient solution to the differential equations. We choose $n_c$ such that the integration time represents at least 2000 periods of both the signal and pump. We take only the last quarter of the DTI solution, which we find is sufficient to remove transient effects from the output signal. To see that the transient solution quickly dissipates, we provide examples of the time-domain output signal computed via DTI for amplifiers with inductive blocks composed of a chain of 10 RF-SQUIDs. These examples, shown in Fig.~\ref{fig:transient}, demonstrates that the transient solution gives way to the steady-state in the long-time regime for amplifiers with inductive blocks with monotonic current-phase relations. For those with nonmonotonic current-phase relations, the integrated solution does not become periodic. Since we only select the last quarter of our DTI solution, and only propose amplifiers with inductive blocks monotonic current-phase relations, we avoid including transient effects in analyzing the output of these amplifiers. In order to extract the power of the amplified signal frequency, we use the DFT with 50 samples of the DTI solution per signal period.

In order to begin the optimization procedure, we must choose initial parameters for the JPA circuit elements and for $\phi_{\tin}$. Initial amplifier designs were based on the RF-SQUID design shown in Fig.~\ref{fig:RFSquidAmp}, for which $L$ and $i_\text{c}$ were chosen in order to design a $\omega_\text{s}=6~(2\pi)~\text{GHz}$ amplifier. Then, the pump amplitude $A_\text{p}$ is chosen such that for small signal amplitude, the gain of the amplifier is 20~dB. Our circuit designs all represent extensions of this RF-SQUID design with additional shunting with Josephson junctions and with applied current bias.

We then apply the optimization procedure described in Section~\ref{sec:numerics} with circuit parameters, and sometimes damping rate $K$, as variables. With this procedure, we find optimal circuit parameters which minimize deviations of the gain from 20~dB, and which consequently improve the PAE.

\section{Power Spectrum of Nondegenerate Amplifier Near Gain Compression Point}\label{appx:spectrum}

\begin{figure}
    \centering
    \includegraphics[width = \textwidth]{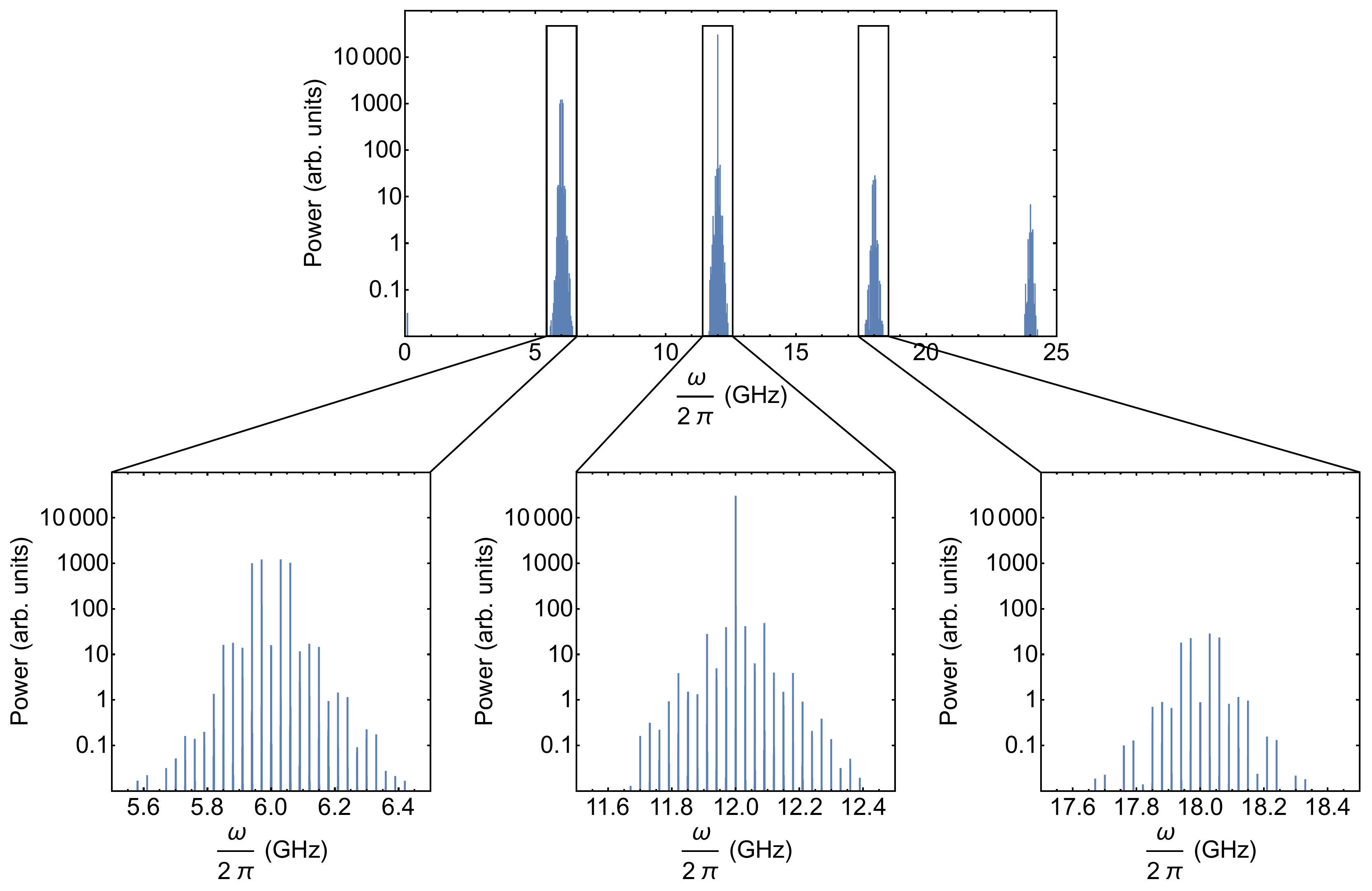}
    \caption{Fourier analysis of the output of the nondegenerate RF-SQUID amplifier near the saturation point. Here we have the two signal frequencies $\omega_1=101/201 \, \omega_\pump$ and $\omega_2=101/200 \, \omega_\pump$, which are given the same input power (-105.9~dBm). The top portion of the figure shows the spectrum of the full band, from 0~GHz to 25~GHz. We provide further detail of the power spectrum around 6~GHz, 12~GHz, and 18~GHz, which shows the vicinity of the signal frequencies and the pump frequency, as well as second order terms which mix pump and signal. The two peaks just above 6~GHz represent the two signals which are approximately 1 to 2 orders of magnitude higher than the next highest products (except their idlers, which are approximately equal), and are lower than only the pump tone at 12~GHz.}
    \label{fig:powerSpectrum}
\end{figure}

Here we investigate the power spectrum of the nondegenerate RF-SQUID amplifier when used to amplify two signals near the saturation point. This serves to identify the behavior of other intermodulation products in addition to the third order terms which are relevant in IP3 analysis. We provide the output power spectrum of the band with focus on the frequencies near the signal and pump tones, as well as $\omega_\text{p}+\omega_\text{s}$, in Fig.~\ref{fig:powerSpectrum}. That is, we can investigate intermodulation products around 6~GHz, 12~GHz, and 18~GHz.

The highest peak in the spectrum is the pump tone at 12~GHz, while the two signals and their idlers on either side of 6~GHz follow. The next largest components, the third order products plotted in Fig.~\ref{fig:kip3} and the second order terms at $\omega_\text{p}+\omega_\text{s}$, remain a factor of at least 21 below the two signal tones. We therefore believe intermodulation products remain manageable up to saturation power, and we note that the saturation point is reached not because of pump power being consumed by these intermodulation products, but because of fundamental limits to the amplifier's ability to amplify at the signal frequency.

\section{Tolerance of amplifier performance to variations in circuit parameters}\label{appx:variations}

The JPA circuit designs provided in this paper and their predicted PAE are derived by considering ideal circuits, where repeating blocks are able to be manufactured identically and with no variation in parameters. We consider here how the performance of the degenerate RF-SQUID amplifier specified in Table~\ref{tab:parameters1} is affected by variations in the fluxes through each of the 10 loops in our circuit. To do this, we simulate the circuit with intermediate nodes and stray capacitances as in Sec.~\ref{sec:modeCouple}. In particular, we simulated the RF-SQUID amplifier circuit with 10 equations of motion, where each equation corresponds to a single RF-SQUID, and the flux through each SQUID's loop is varied. Since this is a computationally difficult problem, we computed amplifier performance for a single set of fluxes which varied from the optimized flux by a random amount selected via a normal distribution. We increased the standard deviation of this distribution to test the effect of increasing variations on amplifier performance.

\begin{figure}
    \centering
    \includegraphics[width = 0.6\textwidth]{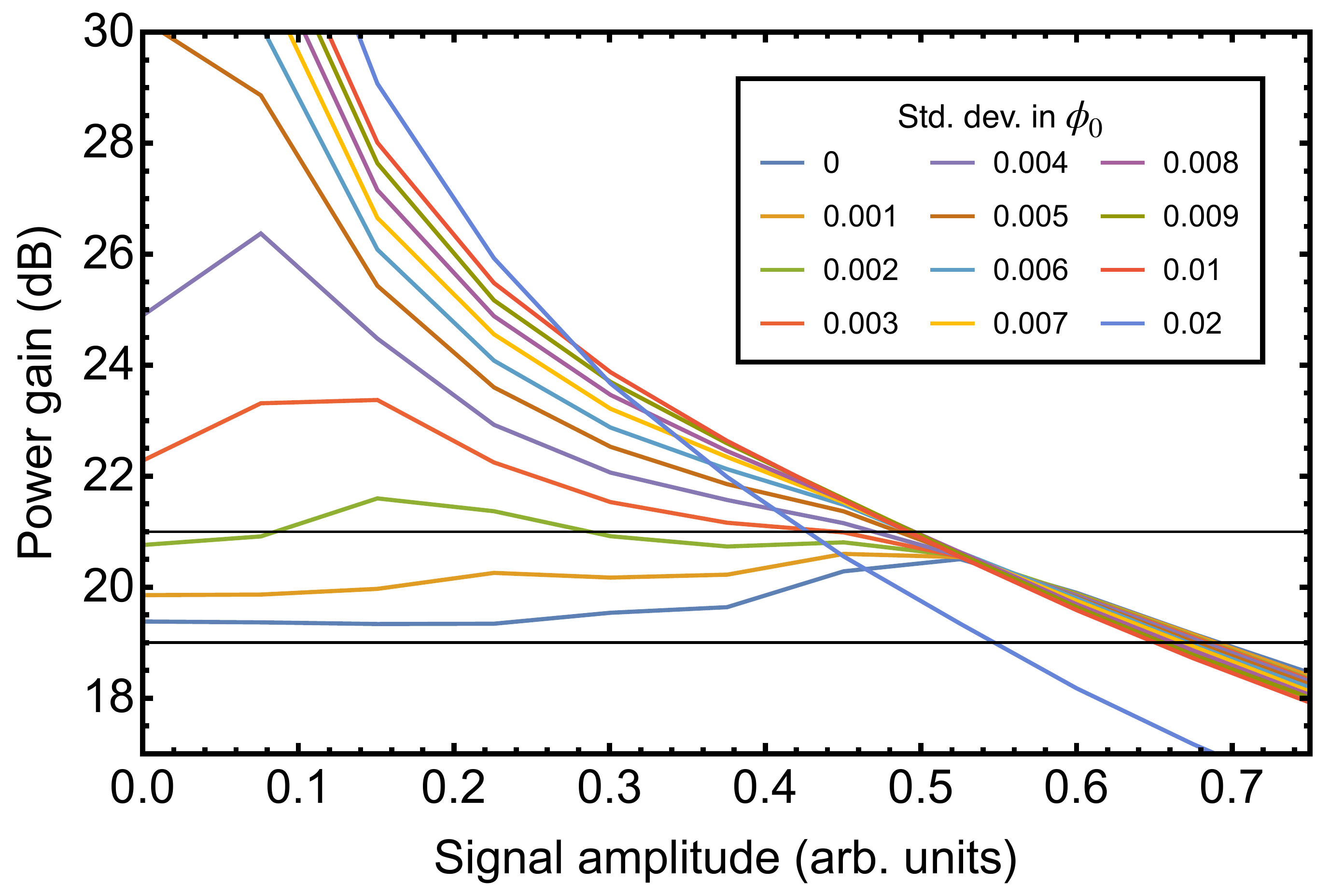}
    \caption{Gain versus signal amplitude for different amounts of variation in the external flux through each RF-SQUID loop. In the low signal amplitude regime, increasing the variation increases gain, leaving the desired $\pm 1~\dB$ region. For higher signal amplitudes, the performance tends to be less sensitive to variations in flux.}
    \label{fig:variations}
\end{figure}

We find that when adding variations to the flux through the loop, we must adjust the pump power to maintain our 20~dB target gain in the low signal regime. In Fig.~\ref{fig:variations}, we show gain versus signal amplitude for varying standard deviation of the fluxes without adjusting pump power. The gain at low signal amplitude quickly leaves the 20~dB $\pm$ 1dB region, as indicated by the black horizontal lines.

\begin{figure}
    \centering
    \includegraphics[width = \textwidth]{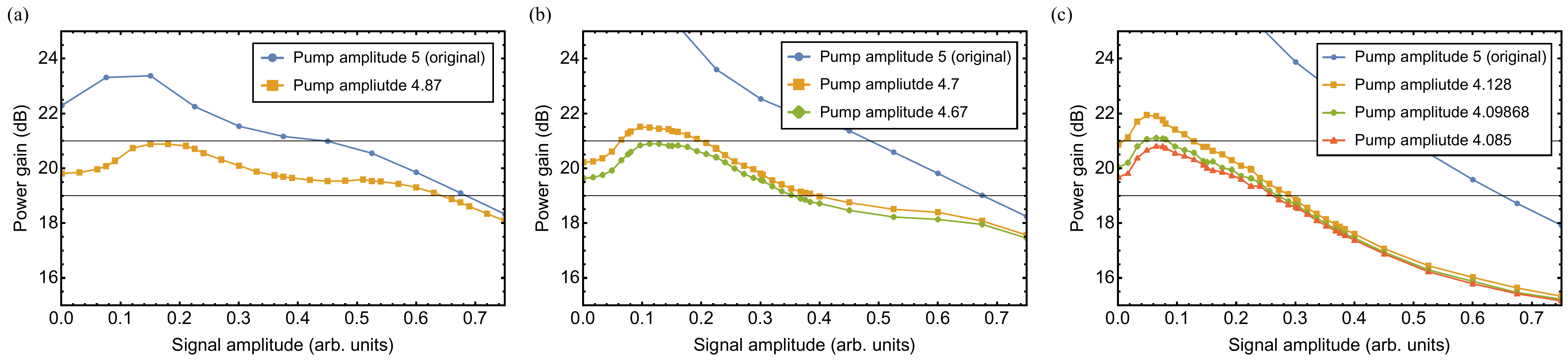}
    \caption{Gain versus signal amplitude for select amounts of variation in external flux. We reduce signal power gradually to re-enter the $\pm 1~\dB$ region. (a) Variations with standard deviation of $0.003 \phi_0$. (b) Variations with standard deviation of $0.005 \phi_0$. (c) Variations with standard deviation of $0.01 \phi_0$.}
    \label{fig:reducePump}
\end{figure}

We then consider decreasing the pump power for different amounts of flux variation to return to our 20~dB $\pm$ 1dB regime and recompute PAE. We find that for a standard deviation of $0.003 \phi_0$, we are able to retain a PAE of at least 34.3\% (Fig.~\ref{fig:reducePump}(a)), as compared to 37.9\% without the added flux variations. That is, with this level of variation in flux, we lose only a marginal amount of efficiency. Increasing the standard deviation to $0.005 \phi_0$, we find a PAE of at least 11.3\% (Fig.~\ref{fig:reducePump}(b)). For a standard deviation of $0.01 \phi_0$, the PAE is at least 8\% (Fig.~\ref{fig:reducePump}(c)). Therefore, for this JPA circuit design, it is ideal to keep variations in threaded flux below $0.003 \phi_0$ to $0.005 \phi_0$.

\begin{figure}
    \centering
    \includegraphics[width = 0.6\textwidth]{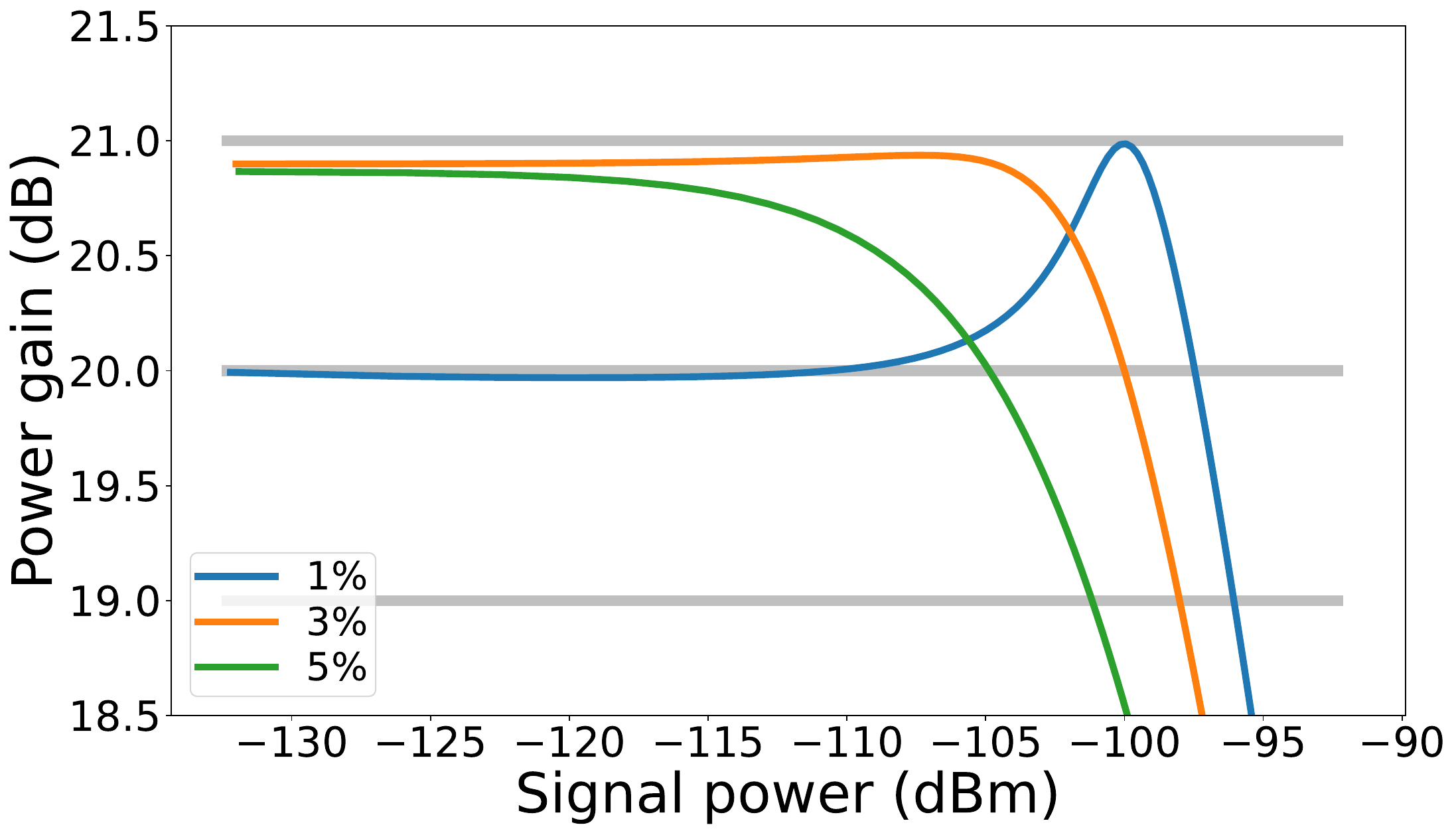}
\caption{Gain versus input signal power for 1\%, 3\%, and 5\% decreases in the linear inductances in the degenerate RF-SQUID amplifier. The pump power and pump/signal frequencies have been adjusted to recover PAE. As compared to the original design, in the 1\% case the pump and signal frequencies are increased by 1.4\% and the pump power is -72.2~dBm. For the 3\% offset, the frequencies are increased by 3.7\% and the pump power is -71.9~dBm. For the 5\% offset, the frequencies are increased by 5\% and the pump power is -71.5~dBm.}
    \label{fig:LOffsets}
\end{figure}

Another possible manufacturing defect is a uniform offset of a device parameter from the ideal value described in this paper. In practical use, the ideal operating point of an amplifier is determined by adjusting both pump power and frequency so as to maximize PAE. In order to demonstrate how this can mitigate uniform manufacturing defects, we consider the case of our degenerate RF-SQUID circuit specified in Table~\ref{tab:parameters1} with a decrease of 1\%, 3\% and 5\% of the linear inductances from their ideal values. We adjust the pump power and signal/pump frequency together in order to achieve the highest possible PAE under these defects, as compared to 37.9\% with no defects. With 1\% offset, we find that the PAE can be restored to at least 33.0\%, with 3\% offset, we can achieve at least 20.8\% PAE, and with 5\% offset, we can achieve at least 9.2\% PAE. We provide the gain versus input signal power in these three cases in Fig.~\ref{fig:LOffsets}. We therefore find that it is possible to recover amplifier performance in practical amplifiers which are subject to the limitations of manufacturing precision.

\end{document}